**SURVEY**  **Open Access**

# On building machine learning pipelines for Android malware detection: a procedural survey of practices, challenges and opportunities

Masoud Mehrabi Koushki[1*], Ibrahim AbuAlhaol[2], Anandharaju Durai Raju[3], Yang Zhou[4], Ronnie Salvador Giagone[1] and Huang Shengqiang[1]

## Abstract

As the smartphone market leader, Android has been a prominent target for malware attacks. The number of malicious applications (apps) identified for it has increased continually over the past decade, creating an immense challenge for all parties involved. For market holders and researchers, in particular, the large number of samples has made manual malware detection unfeasible, leading to an influx of research that investigate Machine Learning (ML) approaches to automate this process. However, while some of the proposed approaches achieve high performance, rapidly evolving Android malware has made them unable to maintain their accuracy over time. This has created a need in the community to conduct further research, and build more flexible ML pipelines. Doing so, however, is currently hindered by a lack of systematic overview of the existing literature, to learn from and improve upon the existing solutions. Existing survey papers often focus only on parts of the ML process (e.g., data collection or model deployment), while omitting other important stages, such as model evaluation and explanation. In this paper, we address this problem with a review of 42 highly-cited papers, spanning a decade of research (from 2011 to 2021). We introduce a novel procedural taxonomy of the published literature, covering how they have used ML algorithms, what features they have engineered, which dimensionality reduction techniques they have employed, what datasets they have employed for training, and what their evaluation and explanation strategies are. Drawing from this taxonomy, we also identify gaps in knowledge and provide ideas for improvement and future work.

## Introduction

Android has become the primary target of mobile malware attacks, due, in no small part, to its high market share (StatCounter 2021). Reports show that the number of new malware samples discovered for Android has been increasing steadily over the past decade. Kaspersky, for example, reported detecting over 5.5 million malicious packages in the year 2020, which was a 62% increase compared to 2019 (Kaspersky 2021). A similar trend was observed by McAfee from 2018 to 2019 when the total number of mobile malware samples increased by over 25% (McAfee 2021).

This large number of samples has rendered manual analysis and classification of them infeasible. And, in turn, has made it essential for the defenders (e.g., cybersecurity researchers, mobile app store holders, or antivirus companies) to try and automate the detection process. This need, in turn, has lead to an influx of research papers that investigate various solutions for automatic detection of Android malware. Enck et al. (2009), for

*Correspondence: mehrabimail@gmail.com
[1] Huawei Technologies Canada Co., Ltd, Vancouver, Canada
Full list of author information is available at the end of the article example, made one of the very first attempts at this. They performed security requirement analysis and created a set of rules to detect malicious intent in Android applications. Using their approach, they were able to detect five new malicious apps on Google play store.[1] Another notable example was the work by Arp et al. (2014) where they proposed DREBIN, a well-known Android malware detection systems which combines program analysis with Machine Learning (ML) to achieve over 93% accuracy.

DREBIN is far from being the only approach that uses ML, however. Investigating the landscape of this research field shows that ML is championed as the most promising approach to achieve accurate malware detection. In fact, while there are a few papers advocating for signature- or pattern-matching-based detection (Tong and Yan 2017; Grace et al. 2012; Talha et al. 2015; Wong and Lie 2016), the majority seem to focus on ML (we counted over 100 of them when gathering literature for this paper, as we discuss in "Taxonomy of android malware detection approaches" section). This is unsurprising, given the promise ML has shown in other domains, such as dynamic intrusion detection (Buczak and Guven 2015) or IoT threat hunting (Raju et al. 2021).

However, the high number of published papers has also made it difficult for new researchers (or practitioners) to easily catch up with the state-of-the-art. This is evidenced, at least partially, by the fact that a high number of published papers use similar feature sets, but fail to clearly distinguish their approach from prior ones (as we discuss in "Gaps in knowledge and future research directions" section). The matter is further complicated by the intricacies of applying ML to a new problem domain, including issues such as the need for representative feature engineering, run-time constraints, usability considerations, and other functional and non-functional requirements (Gift and Deza 2021; Das and Cakmak 2018).

There have been attempts at solving this issue by conducting surveys of the existing literature (Naway and Li 2018; Feizollah et al. 2015; Narudin et al. 2016; Arshad et al. 2016; Ye et al. 2017; Souri and Hosseini 2018; Yan and Yan 2018). For example, Naway and Li (2018) provided a comprehensive review of the use of deep learning for Android malware detection, creating a taxonomy based on features and datasets used. Similarly, Feizollah et al. (2015) provided a review of the feature selection techniques used, providing categorization based on manual or algorithmic selection. Another notable example is the work done by Yan and Yan (2018) where they reviewed the landscape of dynamic mobile malware detection techniques.

None of the current survey papers, however, cover the techniques used in all stages of the ML pipeline–from data collection, to feature extraction, to model use in the end (we discuss these stages in more detail in "Machine learning pipeline" section). Yan and Yan (2018), for example, only focused on dynamic feature extraction, omitting the large body of knowledge that exists on static Android malware analysis. Likewise, Naway and Li (2018) only considered deep learning techniques, leaving out all papers that use simpler (e.g., linear) ML techniques which could be more suitable for deployment on constrained mobile platforms.

In this paper, we address this gap by providing a comprehensive overview of how the literature has used ML end-to-end (from data collection all the way to model explanation). We provide a procedural taxonomy of the proposed solutions, structured based on the stages of a typical ML pipeline (Docs M 2022). Consequently, we discuss (1) how researchers have collected data for training and testing their models, (2) how and what features they have engineered/extracted from Android apps, (3) how they have represented the features for modeling, (4) what dimensionality reduction (feature selection/elimination) techniques they have used, (5) what machine learning algorithms they have tried, (6) what metrics they have used for evaluation of their models (and what the results have been), and, lastly (7) how they have deployed or explained their model. Additionally, in each stage, we identify gaps and provide ideas for improvement and future work.

We should note that the aim of this paper is not to compare (e.g., in terms of the final accuracy) the different approaches taken by the literature. Attempting to do so would be invalid without performing independent measurements of the efficacy of each approach, all while making sure to keep the conditions of the study consistent for all solutions. Instead, we aim at providing a general overview of this research field, to help newcomers catch to speed with the status quo.

In summary, therefore, this paper makes the following contributions to the field of ML-based Android malware detection:

- We provide an overview of the existing literature, based on a review of 42 highly-cited papers published in reputable sources. The review can help new researchers better understand the status quo of the research field.
- We provide a procedural taxonomy of the proposed solutions: Unlike previous survey papers, our taxonomy is structured based on stages of an ML pipe-line. Therefore, it can help new practitioners to better understand how to approach the implementation of each stage in their ML pipelines.
- Drawing from our review, we identify and discuss gaps in knowledge, and provide ideas for future research and improvement.

---

[1] This is the official Android app store operated by Google, available at https://play.google.com.

The rest of this paper is organized as follows: "Background" section provides an overview of the Android operating system architecture, the structure of its application (apps), and its security model. We also introduce the stages of a typical machine learning pipeline and what they entail in the case of Android malware detection. In "Methodology" section, we describe the methodology we used to conduct this survey study. In "Taxonomy of android malware detection approaches" section provides our procedural taxonomy of ML-based Android malware detection, discussing what researchers have tried in each stage of the ML pipeline. In "Gaps in knowledge and future research directions" section we discuss gaps we have identified in the literature and provide ideas for follow-up research. To give more context to our findings, "Reviewed papers" section provides a timeline and brief summaries of all the papers we reviewed. in "Related work" section discusses the related work, which covers existing survey papers on Android malware. And, lastly, "Conclusion" section concludes this paper.

## Background

To aid readers with a better understanding of how the proposed approaches operate, we provide in this section brief overviews of the architecture of Android, the structure of its apps, and its security mechanisms. We also discuss the different stages of a typical ML pipeline which informs the design of our taxonomy.

### Android system architecture

Android is an open-source operating system which consists of a modified Linux kernel and a dedicated software stack designed to create a mobile computing platform (Project AOS 2021). It enables users to install and run first- and third-party applications that perform a variety of tasks, such as taking pictures, location finding and navigation, and internet communications. Figure 1 depicts the architecture of Android and the components included in each of its layers. As seen, its main components include:

- **Linux kernel**: At the heart of Android is a modified Linux kernel which is responsible for communications with hardware, and providing a platform for device drivers (Project AOS 2021). Android's main modifications to the Linux kernel include a new memory management system called Low Memory Killer (which manages memory more aggressively to meet the requirements of a power constraint mobile environment) and a new Inter-Process Communication (IPC) system called Binder (Project AOS 2021).
- **Hardware abstraction layer (HAL)**: This layer provides an standardized interface for the upper layers to access device hardware capabilities, such as camera or audio (Portal AD 2021c). It is responsible for making Android agnostic to lower-level driver implementations (Project AOS 2021).
- **Android runtime**: This component is responsible for interpreting the code and the running of the apps (Project AOS 2021). Android app are often developed in either Java or Kotlin programming language, which are then complied and stored in a specialized format called Dalvik Executable Format (DEX), optimized for minimal memory footprint (Project AOS 2021). The Android Runtime (ART) then spawns virtual machines that interpret and run DEX files[2]. This component also houses Core Libraries—a set of Java libraries that provide most of the functionalities of Java programming to Android app developers (Portal AD 2021c).

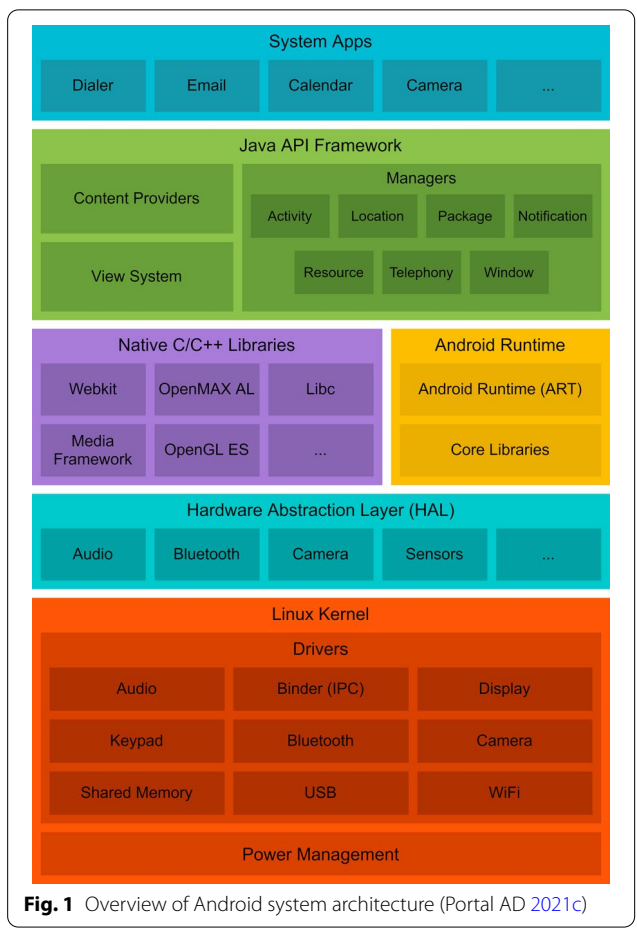

**Fig. 1** Overview of Android system architecture (Portal AD 2021c)

- **Native C/C++ libraries**: These libraries provide access to core system components that are written in C or C++, such as OpenGL graphics APIs (Project AOS 2021). Usually, Android apps access these components through wrappers provided by the Java API Framework (Techotopia 2021). However, apps can also use Android Native Development Kit (NDK) to access the libraries directly (Project AOS 2021).
- **Java API framework**: The framework exposes Java APIs that make the entire feature set of Android available to apps (Portal AD 2021c). Apps can call these APIs to perform various operations, such as accessing storage, drawing UI components, or communicating with other system services to obtain the data they require, such as with notification and location managers.
- **System apps**: In addition to letting users install their own apps, Android ships with a set of pre-installed ones to allow for a ready-out-of-the-box experience (Project AOS 2021). These include, for example, ones for SMS messaging, internet browsing, phone calls and changing the phone's settings.

Malware have been observed to interact with or modify any of the above components, to achieve their objectives. For example, they may use native C/C++ libraries to directly access low level APIs and exploit security vulnerabilities.

**Android app structure**

The way Android apps are structured/written is fundamentally different from traditional desktop (e.g., Windows) applications (Portal AD 2021d). Unlike the latter, Android apps do not have a specific starting point. Rather, they provide a set of independent event handlers that can be called by the OS or other apps (Portal AD 2021d). Apps are built out of 4 fundamental building blocks described below:

- **Activities** represent UI screens that the user can interact with (Inc G 2020). Each one consists of an XML file describing the layout of the UI (e.g., what buttons there are and where they are positioned), and a text file housing the code to handle various related events (e.g., what happens when user clicks on one of the buttons). An app can (and often does) have multiple activities which work together to provide a unified user experience (Inc G 2020). For example, the GMail app may have one activity to show the list of emails in user's inbox, and another activity that presents the "compose email" screen.
- **Services** are components that run in the background for an extended period of time (Inc G 2020). They do not have a user interface, but only provide series of event handlers. An example of a service can be one that plays music in the background, or one that connects to a remote server to fetch latest emails from the users' accounts. Services can be (and often are) started or stopped by other components, such as activities.
- **Broadcast receivers** allow apps to receive and react to signals from Android itself or other apps or components (Inc G 2020). An example of a system-wide broadcast made by Android is BOOT_COMPLETED, which is a signal sent when the phone completes its booting process. An app can subscribe to this broadcast to start some of its components (e.g., the service that checks for new emails) whenever the phone boots. Broadcasts may be sent by both external apps and components with-in the same app.
- **Content providers** allow sharing data between apps or app components (Inc G 2020). They expose an entry point to other apps, through which they can fetch, modify or delete the app's shared data. For example, the Google search app might create a content provider to allow other apps to access Google search results. Access to content providers is through URI addresses.

The above components (whether they are situated within the same app or in different ones) communicate using *Intents*. Put simply, an Intent is a messaging object that is used to request an action from another component (Portal AD 2021a). For example, one Activity can start another by sending an Intent to the OS, specifying the class of that Activity. Intents are also used by broadcast receivers to specify which broadcasts they accept.

Apps can be written in Java, Kotlin and C/C++, which are then transformed to an assembly-like language called *smali* (Portal AD 2021d). Smali codes of each app are stored in DEX format in its "*classes.dex*" file (Portal AD 2021d). This file on its own, however, is not sufficient for running the app. Each app also ships an XML file named "*Manifest.xml*", which includes the following information (Portal AD 2021d):

---
[2] Prior to version 5.0, Android used a different runtime called Dalvik (Project AOS 2021).

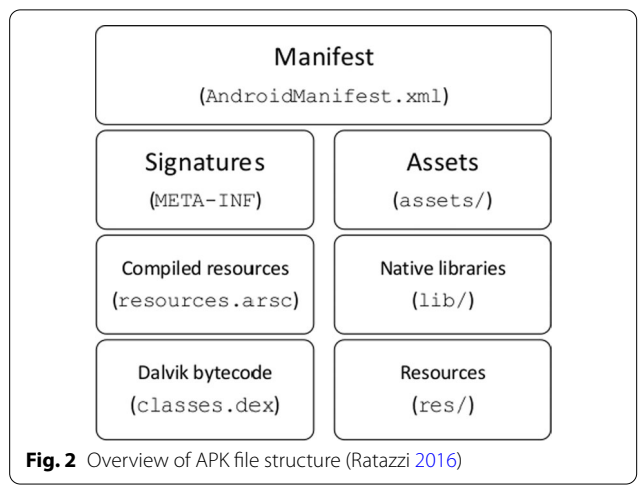

**Fig. 2** Overview of APK file structure (Ratazzi 2016)

- Permissions that it requires, such as access to GPS, or the user's files. The app may also define its own permissions (to be granted to other apps or components). Permissions are explained in more detail later in this section.
- Runtime environment information, including the minimum Android version (API level) that it requires for successful execution.
- Hardware or software features that it requires. For example, an app might require the phone to have a camera, or a fingerprint scanner.
- Libraries that it needs to be linked against (e.g., the Google Play Services).
- A list of components (Activities, Services, Broadcast receivers, or Content providers) that it includes. For each component more information may also be provided. For example, Broadcast receivers may also include Intent filters specifying which broadcasts they accept.

As part of the building process, Android builder tool packages the app's classes.dex file along with its Manifest and any other necessary data and resources (e.g., XML files defining the strings used in the app) into a compressed zip file, with an APK extension. Figure 2 illustrates the structure of an APK file. As seen, the file may also include cryptographic signatures to identify the creator of the app, and native libraries that could be linked to its Java code using Java Native Interface (JNI).

Android malware are often distributed as APK files. Hence, since APKs are zip files, they are easy to extract and examine. Malware analysts usually focus on the Manifest and classes.dex files in each APK. This is because the list of app components and permissions in the Manifest can give clues to the intentions of the app. And, the contents of the DEX file can often be disassembled (sometimes even de-compiled) to investigate the app's behavior. Newer malware samples, however, have been observed to obfuscate the contents of the DEX file to thwart reverse-engineering efforts (Casolare et al. 2021).

**Android security mechanisms**

Android provides several security mechanisms to prevent apps from performing unauthorized actions to harm the device, operating system or the user's data.

Firstly, Android sandboxes apps when executing them. It simply leverages the file-based access control that it has inherited from Linux. Each app runs in its own process and with a unique Linux user ID which limits its access to other apps' data, and to system resources (Portal AD 2021d).

Additionally, Android implements SELinux (Team S 2020)—a Linux kernel extension that provides more fine-grained access control than the file-based model described above. SELinux enables both Mandatory Access Control (MAC) and Role-based Access Control (RBAC) (API A 2020). MAC allows users and objects (e.g., files, documents, etc) to be assigned security levels. Then, a user is only allowed to access an object if they have the necessary security level (Sandhu and Samarati 1994). In RBAC, on the other hand, accesses are not directly assigned to users. Instead, they are associated with roles. To acquire a certain access, users have to be made members of the appropriate role (Osborn et al. 2000). Android's goal for implementing SELinux is to preserve the security of the OS, even if a system service is compromised.

Lastly, Android has an additional access control mechanism called Permissions, which is designed to prevent apps from accessing sensitive system resources without the user's consent. It prevents apps from making certain system calls, unless they are explicitly granted the permission in advance (Elenkov 2014). For example, any app that wants to access the phone's camera has to request the *android.manifest.permission.CAMERA permission*. The user then needs to manually approve each request.

There are more than 250 permissions defined by Android (Inc G 2020) which have different levels of sensitivity: "Normal" permissions are granted automatically to apps without needing user's approval. "Dangerous" permissions, like Camera, are those that require user approval. And, "System" and "Signature" permissions can only be granted to apps that are developed/signed by the OS builder.

The recently introduced Android 12 has put further privacy restrictions on apps (Inc G 2021). For example, users can now choose to only provide approximate location data to apps. And, apps' access to motion sensors are rate-limited. Also, if an app is not used for a few months, all the permissions granted to it will be revoked. Finally, there are also more restrictions on when apps can launch services.

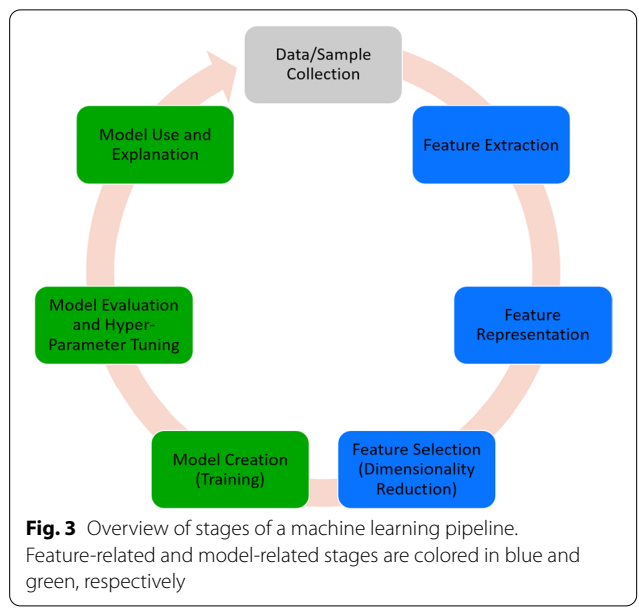

**Fig. 3** Overview of stages of a machine learning pipeline. Feature-related and model-related stages are colored in blue and green, respectively

Android malware have been found to abuse all the above security mechanisms. Permissions in particular have been a heavy focus of research for detecting Android malware, as benign apps tend to request fewer permissions than malicious ones. We will discuss this matter further in "Taxonomy of android malware detection approaches" section.

**Machine learning pipeline**

Machine learning has seen a large surge in use over the past decade. It has shown to be a promising tool to solve long-standing computational challenges, such as image processing (e.g., for medical diagnosis (Fatima and Pasha 2017)), voice recognition (Padmanabhan and Johnson Premkumar 2015), and cybersecurity (Buczak and Guven 2015; Xin et al. 2018).

Generally, after requirement analysis, the use of machine learning for any purpose involves devising a multi-stage pipeline (Docs M 2022; Gift and Deza 2021). It starts by identifying the data required for training the model, then continues by collecting the data and preparing them for digestion by the model (Docs M 2022). Figure 3 provides an overview of the different stages of such pipeline, which include (Das and Cakmak 2018; Docs M 2022; Gift and Deza 2021):

- **Data/sample collection**: This stage involves collecting a representative sample of the data required for training the model. It also includes establishing the ground truth. For Android malware detection, specifically, this entails either finding a readily-available dataset of malware and benign APKs, or creating one through scraping online sources. Establishing ground truth entails labeling the APKs in the dataset as either malware or benign.
- **Feature extraction**: This stage involves extracting from each sample some information that can aptly represent it. This is done to avoid feeding unnecessary and irrelevant data to the model, which may cause performance or efficacy issues. In case of Android malware, doing so usually entails processing the APKs to extract information that could point to malicious or benign intent, such as presence of certain APIs or permissions.
- **Feature representation**: The extracted features are usually not in ideal format for direct digestion by the model. Hence, this stage involves converting them to the appropriate format. For APKs, the features are usually in string format, which is sub-optimal for model training. Thus, various techniques are used to convert them to numbers (e.g., by one-hot encoding) or more complex structures, such as graphs.
- **Dimensionality reduction (feature selection/elimination)**: The number of extracted features are usually too high, which can cause performance issues. To avoid this, feature selection/elimination techniques [e.g., information gain (Alzaylaee et al. 2020)] are used to eliminate less important ones, or only select a subset of features, and improve model efficiency. In case of Android malware, this stage usually involves deciding which parts of APKs to ignore when analyzing them.
- **Model creation (training)**: Usually planned well in-advance, this stage involves devising the architecture of the ML model and then training it using the features obtained in the prior stage. Depending on the objectives, the designers might select a simple model, like linear regression, or a complex one, like a deep neural network. Ensemble models could also be considered. For Android malware detection, various types of models have been designed, as we discuss later in this paper.
- **Model evaluation**: Once the model is trained, it is often tested to ascertain a desired level of performance. This involves using a test dataset to measure various performance metrics (e.g., false positive rate, precision, or recall). It may also include hyper-parameter tuning or model calibration, which is to make sure the probability distribution of the model outcome matches that of the training data. For Android malware detection, the most viral metric to evaluate is the model's accuracy, in terms of its capability to distinguish between malware and benign. However, other aspects may also be evaluated, such as runtime performance or resilience to evasion attacks.
- **Model use**: This last stage involves deploying the model for the intended use. It may also involve devising schemes for explaining the outcomes of it to the end users. In case of Android malware, specifically, this stage usually involves deciding where the model can be deployed (i.e., on the phone or on a server), and could also include schemes for communicating to the end users why an APK is labeled as benign or malware.

Note that the pipeline described above is an iterative one. Accuracy of any trained model may drop over time due to concept drift (e.g., changing distributions) or other external factors, which requires collecting fresh data and/or re-training and re-tuning the model. Higher accuracy requirements might also emerge at any time, which cloud lead to new iterations.

Using ML for Android malware detection requires passing all the above stages. However, we found that no survey paper has provided a comprehensive view of how researchers have approached all of them. This paper fills this gap by providing a taxonomy that covers all the stages.

### Methodology

In this section, we describe the methodology we used for conducting the survey study. Inspired by Naway and Li (2018), our methodology consisted of the following steps:

- **Step 1—literature gathering**: We used keywords "android malware detection", "android malware classification", and "android malware machine learning" to find relevant articles on Google Scholar (Google 2021), ACM Digial Library (ACM 2021) and IEEE Xplore Digital Library (IEEE 2021). We read the abstract of the top 50 search results and composed a list of papers related to our research topic. We excluded papers according to the following criteria:
  – Duplicated papers that we previously found in another library.
  – Papers that used methods other than machine learning (e.g., requirement analysis or sequence matching) for malware detection. A notable example of such papers is MamaDroid (Onwuzurike et al. 2019).
  – Non-peer-reviewed articles (e.g., manuscripts published on ArXiv).
  – Non-English papers.

  Our final list contained 113 papers.

- **Step 2—in-depth review**: We sorted the list of papers (composed in Step 1) by citation counts, which were obtained through Microsoft Academic Search API (Microsoft 2021). We then conducted a detailed review of 42 of them, while prioritizing those that included approaches not already in our taxonomy (a.k.a., those that introduced a new approach for at least one stage of the ML pipeline). We stopped after 42 papers because we reached theoretical saturation and latter reviewed papers did not enrich the taxonomy any further. For each paper, we were interested in obtaining the following information from it:

  – The year of publication: This was to create a chronological history of the publications.
  – The source and size of the dataset used for training or testing the model
  – The features engineered to represent the samples in the dataset as well as the rationale behind the features, and the tools used to extract them.
  – The way the features were represented (e.g., boolean vectors) for digestion by the model
  – The approach that was taken to reduce dimensionality and eliminate/select features
  – The structure of the model
  – The way the model was evaluated, be it based on the confusion-matrix, such as accuracy, precision, recall; or other aspects, such as runtime performance for resilience to attacks. Hyper-parameter tuning and model calibration was also investigated.
  – How the model was deployed (on-device or off-device) and whether there was any attempt towards explanation
  – The novel contributions of the paper (e.g., engineering a new type of feature)

- **Step 3—taxonomy building**: We performed a card-sorting exercise to create an affinity diagram of the approaches proposed by the reviewed papers. The aim was to identify broader themes of android malware detection techniques, and to find potential gaps that can inform future research. The resulting taxonomy, presented in "Taxonomy of android malware detection approaches" section shows, for example, how dynamic analysis has not been researched as well as static analysis, demonstrating potential for future research.

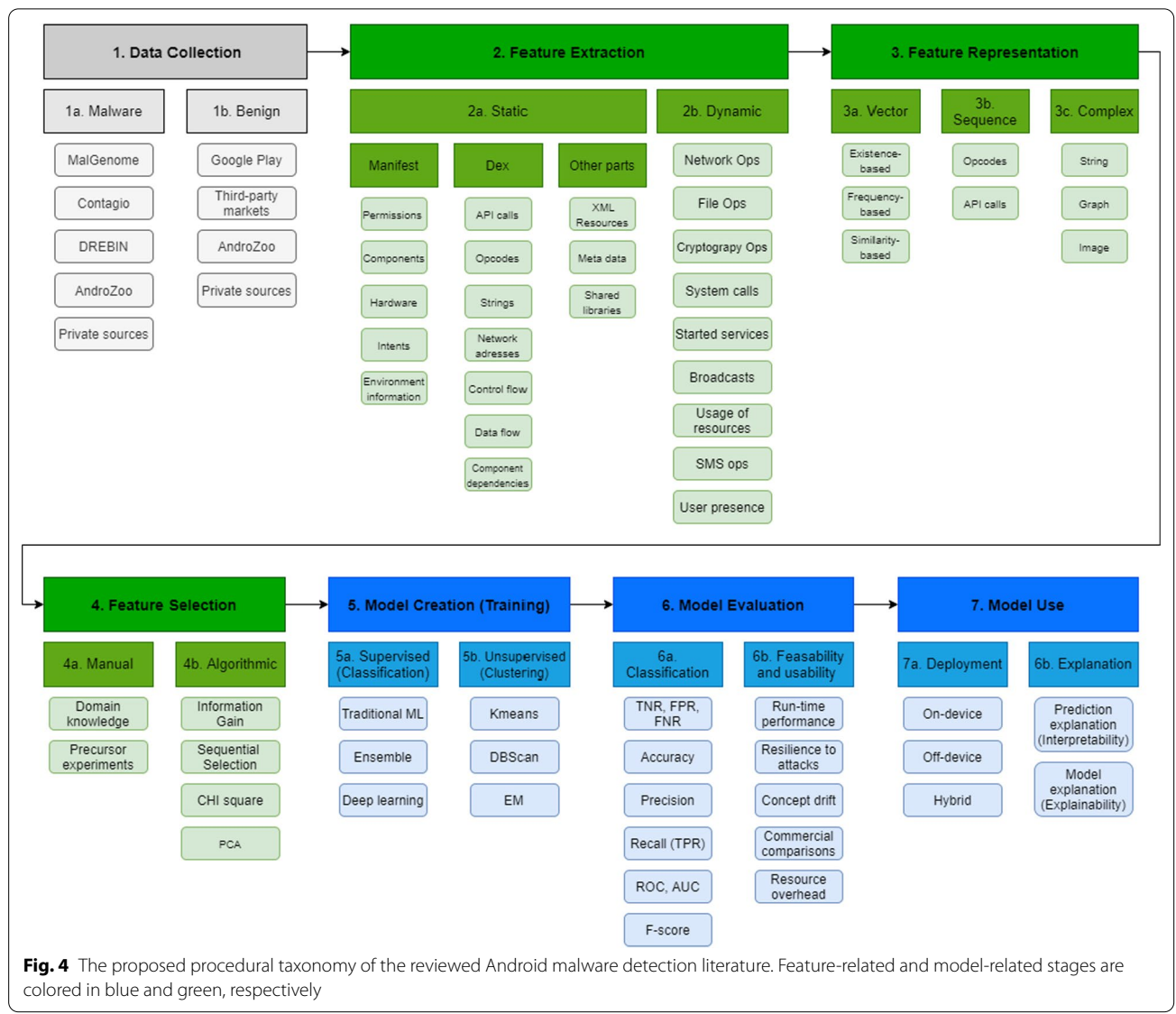

**Fig. 4** The proposed procedural taxonomy of the reviewed Android malware detection literature. Feature-related and model-related stages are colored in blue and green, respectively

## Taxonomy of android malware detection approaches

This section presents our taxonomy of the reviewed papers which is depicted in Fig. 4. As discussed before, the taxonomy is structured based on the stages of an ML pipeline. However, we have also identified different subcategories in each stage, which we will discuss in detail in the following sections.

### Data collection

For ML-based Android malware detection, this stage typically involves collecting a set of APK files. This stems from the fact that the objective of the eventual model is often to simply label APKs as either malware or benign. Also, if supervised learning is being considered, the ground truth needs to be established in this stage as well, meaning each APK in the training set needs to be labeled as either malware or benign. This can be done either by manual analysis or through automated ways, such as using commercial anti-virus products.

Our literature review revealed that researchers have used different approaches for sourcing malware and benign APKs. For malware, as section 1a of Fig. 4 depicts, there are a set of publicly-available datasets [e.g., MalGenome (Zhou and Jiang 2012)] that most papers use (e.g., Liu and Liu 2014; Yuan et al. 2016). For reference of future researchers, we have composed a list of these datasets in Table 1. The table also provides some other information about each dataset, including the number of samples in it, when it was last updated, whether it provides ground truth, and its ease of access. We should note that we only included in our table datasets that provide access to raw APK files. There are datasets that only provide values of extracted features from APKs [e.g., the one by Cai et al. (2020), Cai and Ryder (2017), Cai and Ryder (2020), and Li et al. (2021)]. While these sets might be useful for training certain types of models, we have excluded them as they have limited usefulness for training a general ML pipeline.

**Table 1** Android malware datasets used by the reviewed papers

| Name | Last updated | # APKs | Ground truth | Access | Used by |
|---|---|---|---|---|---|
| Android Malware Genome Project (MalGenome) (Zhou and Jiang 2012) | 2012 | 1260 Malware | Provided | Discontinued | Liu and Liu (2014), Arp et al. (2014), Yuan et al. (2016), Zhang et al. (2014), McLaughlin et al. (2017), Demontis et al. (2019), Yerima (2013), Kim et al. (2019), Tong and Yan (2017), Karbab et al. (2018), Aafer et al. (2013), Peiravian and Zhu (2013), Saracino et al. (2018), Amos et al. (2013), Lindorfer et al. (2015), Suarez-Tangil et al. (2017) |
| DREBIN (Arp et al. 2014) | 2014 | 5560 Malware | Provided | Restricted | Arp et al. (2014), Demontis et al. (2019), Feng et al. (2018), Zhang et al. (2018), Karbab et al. (2018), Suarez-Tangil et al. (2017) |
| M0Droid (Damshenas et al. 2015) | 2015 | 200 Malware | Provided | Restricted | Milosevic et al. (2017) |
| Contagio Mobile (Contagio 2021) | 2018 | ~500 Malware | Not Provided | Open | Yuan et al. (2014, 2016),, Wu et al. (2012), Demontis et al. (2019), Saracino et al. (2018), Lindorfer et al. (2015) |
| VirusShare (2021) | 2019 | 66,727 (Not All Malware) | Not Provided | Restricted | Wang et al. (2014), Kim et al. (2019), Zhu et al. (2018), Xu et al. (2018), Saracino et al. (2018) |
| AndroZoo (Allix et al. 2016) | 2021 | 15,307,857 (Not All Malware) | Provided | Restricted | Feng et al. (2018) |
| VirusTotal (2021) | 2021 | – | Provided | Paid | Sanz et al. (2013), Lindorfer et al. (2015) |
| Private sources | N/A | N/A | N/A | Closed | Liu and Liu (2014), Alzaylaee et al. (2020), Zhang et al. (2014), McLaughlin et al. (2017), Wang et al. (2014), Tong and Yan (2017), Feng et al. (2018), Yerima et al. (2014, 2015),, Wu and Hung (2014), Karbab et al. (2018), Burguera et al. (2011) |

There are also sets that simply provide obfuscated versions of the samples from another set. These can be used for detecting re-packaged malware. The MasVet dataset (Chen et al. 2015) and the dataset provided by Maiorca et al. (2015) are examples of such.

Other than public sources, researchers (e.g., Liu and Liu 2014; McLaughlin et al. 2017; Wang et al. 2014) occasionally use private sets obtained from commercial partners (e.g., McAfee). Lastly, some papers do not disclose the source of their APKs (e.g., Sahs and Khan 2012; Wang et al. 2016), hindering replication of their findings.

When it comes to sourcing benign samples, the most popular approach seems to be scraping Google play store or other third-party markets (e.g., AppChina) for APK files. This approach, however, has the drawback of accidentally mixing malware APKs with benign ones. This is because there is no guarantee that all APKs on official or third-party stores are benign. Evidently, there have been reported cases of malicious apps being listed on such stores (ZDNet 2021). Alternatively, AndroZoo (Allix et al. 2016) has been used occasionally as a source of benign APKs. And, lastly, some researchers use privately-sourced or undisclosed benign sets as well. This division is depicted in section 1b of Fig. 4.

For establishing ground truth, the most popular approach in the literature seems to be submitting the APKs to VirusTotal (2021), and using a threshold on the number of positive detections. For example, one might decide to label any APK that is detected by more than 3 AntiVirus (AV) products on VirusTotal as malware. Aside from using VirusTotal, some papers assume ground truth based on the source of APK. For example, they label any APK they download from Google play store as benign because they assume the store is malware free. A few papers (e.g., Burguera et al. 2011) used other methods, such as using self-written malware or performing manual analysis.

**Feature extraction**

In our literature review, we observed that engineering new features is often positioned as a major contribution of the reviewed works. This suggests the high importance of feature engineering for successful Android malware detection. This is unsurprising given that APKs are large and complex files, containing lots of information (e.g., the graphics of the app) that are irrelevant for malware detection. This makes APKs sub-optimal for direct digestion by ML algorithms, as doing so would create needless processing and storage overhead. We found, as demonstrated in section 2 of Fig. 4, that the literature has tried the two following general ways of extracting representative features from APKs.

*Static analysis*

Static analysis involves extracting features from APKs without executing them, either on a real device or an emulator. Instead, the APK is unpacked and the files with-in it are examined for clues about the app's intentions. Examples include presence of certain suspicious API calls in the bytecode, or usage of requesting permissions in the Manifest file.

Our literature review revealed that researchers have extracted static features from nearly every part of the APK structure we presented in "Background" section (see Fig. 2). This is clearly depicted in section 2a of Fig. 4 as well. Generally, we found that while most papers have focused on the manifest and classes.dex file (e.g., Liu and Liu 2014; Arp et al. 2014; Yuan et al. 2016), occasionally researchers have looked into the resources folder (like Xu et al. 2018), as well as signatures and native libraries (e.g., Lindorfer et al. 2015).

For future researchers' reference, we have composed a list of these static features, in Table 2. For each feature, we provide which part of the APK structure it is extracted from, what is the rationale behind extracting it (according to the authors of the reviewed papers), and, if clarified by the authors, the tool(s) that have been used for extracting the features.

As the table shows, permissions and API calls have been by far the most popular targets for feature engineering. This is unsurprising given that they are most likely to reveal malicious intent. For instance, if a malware wants to spy on the user's location, it needs to request the location permission and then call necessary APIs to obtain the GPS data. Both of these acts can be revealed by an examination of the malware's Manifest and DEX files.

It is, however, interesting that other less obvious data, such as environment info, are also found to be useful for malware detection. For example, Saracino et al. (2018) used data from the market listings of apps. They extract the ratings of each app, its market name, the name of its developer and its number of downloads. They use this data alongside other features, such as requested permissions.

Lastly, we should remark that while some papers extract the same features, they might use them in different ways. Some papers, for example, use requested permissions directly as boolean features (e.g., Arp et al. 2014), whereas others might try combinations of them, such as Liu and Liu (2014) who used pairs of permissions. Similarly, whereas some authors directly use presence of API calls as boolean features (e.g., Yuan et al. 2016), others try to capture the temporal relationship of API calls as well. Hence, they use control or data flow graphs (e.g., Gascon et al. 2013). This overall diversity of feature representation is depicted in the DEX subsection of section 2a of Fig. 4.

*Dynamic analysis*

Dynamic analysis involves running an APK, either on a real device or on an emulator, to observe its behaviors at run time (Yan and Yan 2018). It is usually much more resource intensive than static analysis, as it requires access to an isolated and instrumented execution environment. However, it also has the potential to reveal data not obtainable statically, due to code obfuscation or encryption.

Our literature review revealed that authors have extracted a variety of dynamic features from APKs, which range from kernel-level system calls, to network operations, and SMSs sent. Section 2b of Fig. 4 illustrates a list of such features. Overall, however, the diversity of dynamic features, and the tools used to extract them, is not to the extent we observed with the static ones. We found that most papers (e.g., Yuan et al. 2014, 2016; Alzaylaee et al. 2020; Feng et al. 2018; Wu and Hung 2014; Lindorfer et al. 2015) that perform dynamic analysis use the DroidBox (Project D 2021a) tool, which records the following information about an app:

- **Cryptographic operations**: Whether the app has made any calls to Android's cryptograpic APIs.
- **Network operations**: Any data that is sent/received over the network
- **File operations**: Details of any file reads or writes that the app has performed

**Table 2** Overview of static features extracted from Android APKs by the reviewed papers

| Feature | APK part | Rationale | Tools | Used by |
|---|---|---|---|---|
| Requested permissions | Manifest | Malware tend to request more permissions, and more dangerous ones. | APKTool (2021), aapt (Portal AD 2021b), androguard (Project A 2021) | Liu and Liu (2014), Arp et al. (2014), Yuan et al. (2014, 2016), Alzaylaee et al. (2020), Sahs and Khan (2012), Li et al. (2018), Wang et al. (2014), Wu et al. (2012), Demontis et al. (2019), Yerima (2013), Kim et al. (2019), Zhu et al. (2018), Zhang et al. (2018), Yerima et al. (2014, 2015), Wang et al. (2016), Aafer et al. (2013), Peiravian and Zhu (2013), Saracino et al. (2018), Sanz et al. (2013), Zarni Aung (2013), Lindorfer et al. (2015), Suarez-Tangil et al. (2017) |
| Used permissions | DEX | All requested permissions might not be used. Unused permissions introduce noise and should be eliminated. | APKTool (2021), PScout (Project P 2021b), baksmali (Project B 2021) | Liu and Liu (2014), Arp et al. (2014), Demontis et al. (2019), Lindorfer et al. (2015) |
| Hardware requirements | Manifest | Malware tend to request more sensitive hardware (e.g., Camera) | aapt (Portal AD 2021b) | Arp et al. (2014), Demontis et al. (2019), Sanz et al. (2013) |
| Names and types of app components | Manifest | To detect code reuse (common services, broadcast receivers, or other app components) by malware | – | Arp et al. (2014), Wu et al. (2012), Demontis et al. (2019), Kim et al. (2019), Suarez-Tangil et al. (2017) |
| Filtered intents | Manifest | Malware tend to subscribe to sensitive system broadcasts, such as BOOT_COMPLETE. | – | Arp et al. (2014), Wu et al. (2012), Demontis et al. (2019), Zhu et al. (2018), Lindorfer et al. (2015), Suarez-Tangil et al. (2017) |
| API calls | DEX | Malware may call sensitive or suspicious APIs, such as ones to access SMS. | baksmali (Project B 2021), soot (Project S 2021), androguard (Project A 2021), dexdump (Man Pages U 2021) | Arp et al. (2014), Yuan et al. (2016), Sahs and Khan (2012), Wu et al. (2012), Yuan et al. (2014), Demontis et al. (2019), Yerima (2013), Kim et al. (2019), Zhu et al. (2018), Zhang et al. (2018), Yerima et al. (2014, 2015), Karbab et al. (2018), Aafer et al. (2013), Peiravian and Zhu (2013), Gascon et al. (2013), Yang et al. (2014), Suarez-Tangil et al. (2017) |
| Network addresses | DEX | Malware may commonly communicate with untrustworthy internet hosts. | – | Arp et al. (2014), Demontis et al. (2019) |
| Opcodes | DEX, Shared libraries | Certain sequences of opcodes may reveal malicious intents in apps. | baksmali (Project B 2021), IDA Pro (Hex-rays 2021) | McLaughlin et al. (2017), Kim et al. (2019) |
| Bytecodes | DEX | Certain bytecode sequences may reveal malicious intents in apps. | – | Grace et al. (2012), Xu et al. (2018), Bakour and Ünver (2021) |
| Decompiled Java code | DEX | Certain patterns of code may reveal malicious intent. | dex2jar (Project D 2021b), Procyon (Project P 2021a) | Milosevic et al. (2017), Wang et al. (2016) |
| Linux command strings | DEX & Resources | Malware may use dangerous commands to exploit the phone and gain privileged access. | – | Yerima (2013), Yerima et al. (2014, 2015) |
| Use of encryption routines | DEX | Malware may use encryption to hide their intent. | – | Yerima (2013), Lindorfer et al. (2015), Suarez-Tangil et al. (2017) |

**Table 2** (continued)

| Feature | APK part | Rationale | Tools | Used by |
|---|---|---|---|---|
| Presence of secondary APK or shell scripts | Assets | Malware may hide APK files which will be installed after infection. Shell scripts might be used for exploitation. | – | Yerima (2013), Lindorfer et al. (2015) |
| Environmental Information | Manifest | Malware may target a specific vulnerable execution environment (e.g., Android version). | – | Kim et al. (2019), Suarez-Tangil et al. (2017) |
| Constant strings | Resources | Malware may contain suspicious strings (e.g., fake ads) | APKTool (2021) | Kim et al. (2019), Zhang et al. (2018), Xu et al. (2018), Suarez-Tangil et al. (2017) |
| Use of Java reflection | DEX | Malware may use reflection to dynamically load code and thwart static analysis efforts. | – | Lindorfer et al. (2015), Suarez-Tangil et al. (2017) |
| Signing certificate data | META-INF | The fingerprint, serial number, owner or other data from the certificate may correspond to known malware authors. | – | Lindorfer et al. (2015), Suarez-Tangil et al. (2017) |
| Presence of native executables or libraries | Lib | Malware often use native code to perform exploits or make reverse-engineering harder. | – | Lindorfer et al. (2015), Suarez-Tangil et al. (2017) |

- **DEX class-load**: Details of classes that the app has loaded dynamically through dexload
- **Information leaks**: Whether the app shows signs of information leak, which is defined as data obtained from sensitive APIs being sent off-device (e.g. over network)
- **Sent SMS**: Details of any SMSs the app sends
- **Phone calls**: Details of any phone calls initiated by the app
- **Services started**: Names of service components started by the app
- **Broadcast receivers**: Whether and what broadcasts are dynamically registered by the app

A few papers (i.e., Vidal et al. 2018; Feng et al. 2018; Burguera et al. 2011; Afonso et al. 2015) also use strace (Man Pages L 2021), a built-in Linux tool, in addition or instead of DroidBox. This allows them to capture system calls at the kernel level, which the authors can then use to detect presence of certain dangerous calls. Alternatively, some authors use custom-developed tools, such as kernel extensions [done by Tong and Yan (2017), Dini et al. (2012) and Saracino et al. (2018)] or apps [done by Saracino et al. (2018) Dini et al. (2012)]. For example, the tool developed by Saracino et al. (2018) can capture SMSs sent, calls made and whether the screen is on, which will indicate the user's presence. Other such custom tools include DynaLog [used by Alzaylaee et al. (2020)] and APIMonitor [used by Wu and Hung (2014)]

Other than DroidBox features and API calls, we observed three noteworthy dynamic features used by the literature. Firstly, Amos et al. (2013) leveraged usage of resources (e.g., battery) to detect malware. The rationale was that malware might use more energy, due to the nature of work it accomplishes. Secondly, Shabtai et al. (2014) focused on network operations to detect maliciousness. They utilize deviations from an app's "normal" network behavior, as a sign of anomaly that can reveal malware. Lastly, Cai et al. (2018) made novel use of the analysis of the Inter-Component Communication (ICC) Intents of apps, in addition to API and system calls, as a measure for detecting malware.

**Feature representation**

Once features are extracted from APK files, they can be represented in different ways, for digestion by the ML models. Examples can be boolean vectors, similarity scores, and graphs. As Table 3 shows, the surveyed papers have used a variety of ways to represent features.

Firstly, the majority of papers have utilized a vector-based representation. In this approach, each APK is represented as a one-dimensional array of values, and any temporal or spacial relationships are ignored (e.g. only the names of the called APIs are considered, but the order by which the calls are made is ignored). As section 3a of Fig. 4 shows, the surveyed work has used different ways of creating feature vectors:

- **Existence-based**: In this simplistic approach, the feature vector consists of a series of True/False values which indicate whether or not the corresponding feature is present in the APK file. For example, in permission-based solutions such as Liu and Liu (2014) and Alzaylaee et al. (2020), each element of the feature vector is mapped to an Android permission, and a corresponding True/False value would mean that the APK has(not) requested that permission. Similarly, in Arp et al. (2014), a vector element indicate whether the APK contains a certain network address, or a suspicious API call.
- **Frequency-based**: Feature vectors can also be frequency-based where each entry is an integer representing the frequency by which a feature has been observed in the APK. For example, in Burguera et al. (2011)'s approach, each element represents how many times the APK has called a certain API. In Milosevic et al. (2017) also, each element represents how many times the corresponding keyword has been detected in the decompiled Java code of the app. Note that, in contrast to the previous type, each element here also demonstrates the prevalence of each feature.
- **Similarity-based**: Lastly, vectors may also be created based on similarities among APKs. In this case, each element of the feature vector represents a decimal number which indicates how similar the current APK is to another one in the dataset. Zhang *et al.*'s approach (Zhang et al. 2014) is an example of this case, where each feature vector element indicates how similar this APK's control flow graph is to that of the other APKs in their dataset. Kim et al. (2019) also uses as features similarities between APKs in terms of Opcode and API call frequency. Lastly, Xu et al. (2018) devised a more complex similarity-based approach. The authors extract bytecodes from the APK. However, instead of representing them as simple boolean vectors, they use a semantic-aware approach called Bytecode2vec. Bytecode2vec codes functionally-similar Bytecodes closer to each other in the feature space. This means that if two bytecodes are semantically similar (e.g., they both do file operations), their corresponding byte2vec values are closer. This allows the approach to be less sensitive to the exact names of APIs used, and rather be able to detect semantically-malicious behavior.

**Table 3** Comparison of ML approaches proposed by the surveyed papers, in terms of feature extraction and representation; model creation, deployment and use; and benefits and drawbacks

| | Feature | | | Model | | | | Contributions |
|---|---|---|---|---|---|---|---|---|
| | Extraction | Representation | Selection | ML Approach | Evaluation | Deployment | Explanation | |
| Liu and Liu (2014) | Static | Boolean vector | Manual | Decision Tree | TPR = 0.813<br>FPR = 0.0046<br>Precision = 0.89<br>Accuracy = 0.98 | Off-device | None | Using "Used" permissions, instead of requested ones can reduce feature noise. Using a two-step detection process can increase performance |
| Arp et al. (2014) | Static | Boolean vector | Manual | SVM | Recall = 0.94<br>Accuracy = 0.93<br>FPR = 0.01<br>Run-time | Hybrid (training off-device, feature extraction and detection on-device) | Uses feature weights to explain predictions | Emphasis on prediction explanation provides clarity to the user, increasing usability. Diversity of features used can alleviate concept drift |
| Yuan et al. (2016) | Static & Dynamic | Boolean vector | Manual | Deep Belief Network (DBN) | Precision = 0.94<br>Accuracy = 0.93<br>FPR = 0.01 | Hybrid (training is off-device, feature extraction and detection is on-device) | Uses feature weights to explain predictions | Using deep learning for Android malware shows promise. Results show that there is resistance to re-packaged malware |
| Alzaylaee et al. (2020) | Static & Dynamic | Boolean vector | Information Gain | Multilayer perceptron (MLP) | TPR = 0.98<br>TNR = 0.91<br>FPR = 0.09<br>FNR = 0.02<br>Accuracy = 0.95<br>F = 0.96<br>AUC = 0.99<br>Run-time | Off-device | None | Using stateful input generation for dynamic analysis has improved code coverage, when compared to other works. Clear run-time performance evaluation is conducted and reported |
| Zhang et al. (2014) | Static | API graph similarity scores | Manual | Naive Bayes | FNR = 0.02<br>FPR = 0.05<br>Recall = 0.93<br>Run-time | Off-device | None | Using semantically-aware dependency graphs lessens the reliance on syntax, helping to detect zero-day malware and potentially alleviating concept drift |

**Table 3** (continued)

| | Feature | | | Model | | | | Contributions |
|---|---|---|---|---|---|---|---|---|
| | Extraction | Representation | Selection | ML Approach | Evaluation | Deployment | Explanation | |
| McLaughlin et al. (2017) | Static | Opcode sequences | None | Convolutional Neural Network (CNN) | Accuracy = 0.87 Precision = 0.87 Recall = 0.85 F = 0.86 Run-time | Off-device | None | Using opcodes and deep learning eliminates the need for manual feature engineering, and also could alleviate concept drift. Thorough run-time performance evaluation is conducted and reported |
| Li et al. (2018) | Static | Boolean vector | SFS | SVM | Accuracy = 0.95 Precision = 0.97 Recall = 0.93 FPR = 2.36 FM = 0.95 | Hybrid (feature extraction on-device, training and detection off-device) | None | Using only "significant" permissions reduces feature noise and model complexity, potentially leading to better accuracy and lower over-fitting |
| Wang et al. (2014) | Static | Boolean vector | MI, SFS, Manual | SVM, Decision Tree, Random Forest | Accuracy = 0.95 TPR = 0.94 FPR = 0.006 F = 0.90 ROC | Off-device | Decision tree rules to explain predictions | The permission ranking used can lead to reduced feature noise and improved accuracy. The model explanation approach is novel and can inspire future efforts |
| Yuan et al. (2014) | Static & Dynamic | Boolean vector | Manual | DBN | Accuracy = 0.96 | Off-device | None | Novel use of deep learning leads to improved accuracy. Using both static and dynamic features can alleviate susceptibility to evasion attacks |
| Wu et al. (2012) | Static | Boolean vector | Manual | K-Means, EM, kNN, Naïve Bayes | Accuracy = 0.93 Recall = 0.87 Precision = 0.96 F = 0.91 | Off-device | None | The performed malware family detection can help human analysts. Classification is augmented with clustering for more accurate detection |

**Table 3** (continued)

| | Feature | | | Model | | | |
|---|---|---|---|---|---|---|---|
| | Extraction | Representation | Selection | ML Approach | Evaluation | Deployment | Explanation | Contributions |
| Milosevic et al. (2017) | Static | Boolean & Integer vectors | None | K-Means, EM; Ensemble of SVM, Naïve Bayes, Decision Tree | Precision = 0.89 Recall = 0.89 F = 0.89 Run-time | Off-device | None | The performed clustering can help with obtaining ground truth for unlabeled samples, based on their neighbors. It can also help with malware family detection and manual analysis |
| Demontis et al. (2019) | Static | Boolean vector | Manual | Secure SVM | Attack-resistance ROC | Hybrid (training off-device, feature extraction and detection on-device) | Uses feature weights to explain predictions | The proposed uniformed feature weights lessens SVM's reliance on any single feature, alleviating certain evasion attacks. Extensive attack evaluations is performed |
| Yerima (2013) | Static | Boolean vector | Mutual information | Bayesian Classification | Accuracy = 0.92 FPR = 0.63 TPR = 0.90 FNR = 0.94 AUC = 0.97 | Off-device | None | The use of Bayesian model makes integrating expert knowledge easier. AUC is provided which allows for easier model comparison |
| Kim et al. (2019) | Static | Boolean vector, Similarity scores | Topological Data Analysis | Deep learning | Accuracy = 0.98 Recall = 0.99 Precision = 0.98 F = 0.99 Resilience to obfuscation attacks | Off-device | None | The great variety of static features used can improve detection accuracy. As does the use of deep learning. A thorough investigation of resilience to different types of attacks is performed and reported |

**Table 3** (continued)

| | Feature | | | Model | | | | |
|---|---|---|---|---|---|---|---|---|
| | Extraction | Representation | Selection | ML Approach | Evaluation | Deployment | Explanation | Contributions |
| Sahs and Khan (2012) | Static | Boolean vectors, Graphs | None | SVM | TPR Precision Recall F-graph | Off-device | None | The novel use of SVM kernels to represent graphs and strings can be inspirational for future work. It can also improve accuracy and alleviate concept drift |
| Feng et al. (2018) | Dynamic | Boolean vector | Chi-square | Ensemble: Stacking of SVM, Decision Tree, Extra Trees, Random Forest, Boosted Tree | Accuracy = 0.97 Precision = 0.95 TPR = 0.97 FPR = 0.016 AUC = 0.97 | Off-device | None | Provides novel insight into the use of ensembles for Android malware detection. A comparison of different ensembling approaches is also provided, showing advantage for stacking. Provides evidence for the unsuitability of kNN for Android malware detection |
| Zhu et al. (2018) | Static | Boolean vector, PCA | TF-IDF | Rotation Forest | Sensitivity = 0.88 Precision = 0.88 Accuracy = 0.88 AUC = 0.89 | Off-device | None | Use of Rotation Forest for Android malware detection can improve accuracy over individual models. However, there might be a performance penalty |
| Zhang et al. (2018) | Static | Boolean vector | None | CNN | Precision = 0.96 Recall = 0.98 Accuracy = 0.97 F = 0.97 Run-time | Off-device | None | The use of a complex neural network architecture like CNN leads to improved accuracy and help with Zero-day malware detection |

Table 3 (continued)

| | Feature | | | Model | | | |
|---|---|---|---|---|---|---|---|
| | Extraction | Representation | Selection | ML Approach | Evaluation | Deployment | Explanation | Contributions |
| Yerima et al. (2015) | Static | Boolean vector | Manual | Random Forest | TPR = 0.97<br>TNR = 0.97<br>FPR = 0.02<br>Accuracy = 0.97<br>Error rate = 0.02<br>AUC = 0.99 | Off-device | None | Use of ensembles can help with detection of Zero-day malware. Features are extracted from both Manifest and DEX, increasing their diversity and potentially alleviating concept drift |
| Yerima et al. (2014) | Static | Boolean vector | Manual | Ensemble: Decision Tree, Logistic Regression (LR), Naïve Bayes (NB) | TPR = 0.97<br>TNR = 0.97<br>FPR = 0.03<br>FNR = 0.02<br>Accuracy = 0.97<br>AUC = 0.95 | Off-device | None | A thorough investigation of the effectiveness of different ensembling techniques is performed. Ensembling can also improve zero-day detection due to model diversity |
| Xu et al. (2018) | Static | Boolean and Bytecode vectors | None | MLP | Accuracy = 0.97<br>TPR = 0.97<br>FPR = 0.02<br>Run-time | Off-device | None | The two-layered detection design can improve performance without loosing accuracy. Use of deep learning reduces the need for manual feature engineering. An investigation of the resilience of the model against different attacks is reported |
| Wu and Hung (2014) | Dynamic | Boolean vector, 2-grams | Manual | SVM | Accuracy = 0.86<br>F = 0.85<br>Recall = 0.82<br>Precision = 0.9<br>FPR = 0.1<br>FNR = 0.18 | Off-device | None | Uses APE, a complex input generation scheme for dynamic analysis, as opposed to simplistic random models used by prior literature. This can improve code coverage and detection accuracy |

**Table 3** (continued)

| | Feature | | | Model | | | |
|---|---|---|---|---|---|---|---|
| | Extraction | Representation | Selection | ML Approach | Evaluation | Deployment | Explanation | Contributions |
| Wang et al. (2016) | Static | Boolean vector | None | DBN | Precision = 0.93 Recall = 0.94 F = 0.93 | Off-device | None | Use of deep learning can improve detection accuracy and eliminate the need for manual feature engineering |
| Karbab et al. (2018) | Static | Vector sequence of API calls | None | CNN | F = 0.96 Precision = 0.96 Recall = 0.96 FPR = 0.031 Family detection Concept drift Attack resilience Run-time | Hybrid (feature extraction on-device; training and detection off-device) | None | Provides a thorough requirement analysis for Android malware detection, which clearly lays out expectations from such system. This allows for better comparison of different solutions proposed by literature. Also, all API calls are considered for analysis, not just a subset, as done by prior work |
| Aafer et al. (2013) | Static | Boolean vector | Manual | Decision Tree | Accuracy ~ 99 TPR ~ 97 TNR ~ 100 Run-time | Off-device | None | Provides a novel way of extracting API calls from DEX files. High run-time performance which leads to increased practicality |
| Burguera et al. (2011) | Dynamic | Integer vector | Manual | K-Means | Detection rate = 0.85 ~ 1.0 | Hybrid (feature extraction on-device; training and detection off-device) | None | Proposed an approach, which compares execution traces of different versions of an app, to detect re-packaged malware (e.g., Trojans) |
| Dini et al. (2012) | Dynamic | Integer vector | Manual | kNN | FPR = 0.001 Family detection Run-time | On-device | None | The approach makes novel use of on-device dynamic analysis for anomaly-based Android malware detection |

**Table 3** (continued)

| | Feature | | | Model | | | |
| --- | --- | --- | --- | --- | --- | --- | --- |
| | Extraction | Representation | Selection | ML Approach | Evaluation | Deployment | Explanation | Contributions |
| Peiravian and Zhu (2013) | Static | Boolean vector | None | Ensemble: Bagging with SVM and Decision Tree | Accuracy = 0.96 Precision = 0.95 Recall = 0.94 AUC = 0.96 | Off-device | None | Provides comparison of use of permissions and API calls for malware detection. Ensemble learning can improve zero-day detection |
| Gascon et al. (2013) | Static | Graph (Integer vector) | None | SVM | FPR = 0.01 Detection rate = 0.89 ROC | Off-device | Using feature weights to explain predictions | Proposes a new way of labeling Dalvik functions for easier call graph generation. Makes novel use of kernels for embedding call graphs for digestion by ML models |
| Saracino et al. (2018) | Static & Dynamic | Integer vector | Manual | kNN | Accuracy = 0.96 FPR = 0.00001 Run-time Battery | On-device | None | Uses a combination of on-device dynamic ML-based detection and signature-based techniques to achieve higher accuracy. Also uses metadata from market listings as features |
| Sanz et al. (2013) | Static | Boolean vector | None | LR, NB, BayesNet, Decision Tree, Random Forest | TPR = 0.91 FPR = 0.19 AUC = 0.92 Accuracy = 0.86 ROC | Off-device | None | A pioneering work in the use of permissions for Android malware detection |
| Zarni Aung (2013) | Static | Boolean vector | Information gain | K-Means, Decision Tree, Random Forest, CART | TPR = 0.97 FPR = 0.15 Precision = 0.84 Recall = 0.97 ROC Area = 0.87 | Off-device | None | Pioneering work in the use of clustering with permissions as features for Android malware detection. Makes novel use of hardware features to detect certain types of malware (e.g., those who use Camera or microphone for spying) |

**Table 3** (continued)

| | Feature | | | | Model | | | | Contributions |
|---|---|---|---|---|---|---|---|---|---|
| | Extraction | Representation | Selection | ML Approach | Evaluation | Deployment | Explanation | | |
| Yang et al. (2014) | Static | Boolean vector | Manual | NB, SVM, Decision Tree, Random Forest | Accuracy = 0.95 FPR = 0.4 Family detection Run-time | Off-device | None | | Novel use of behavioral graph for detecting "malicious behavior" in Android apps, as opposed to simply label them as malware or benign. Clustering is performed to detect malware families |
| Amos et al. (2013) | Dynamic | Boolean vector | None | Random Forest, NB, MLP, BayesNet, LR, Decision Tree | Accuracy = 0.91 TPR = 0.97 FPR = 0.31 Run-time performance | Hybrid (feature extraction on-device; training and detection off-device) | None | | Proposes a distributed system for large-scale detection of Android malware. Dynamic analysis alleviates evasion by code obfuscation or dynamic loading |
| Lindorfer et al. (2015) | Static & Dynamic | Boolean vector | Fisher score | LR, SVM | Accuracy = 0.99 Recall = 0.98 Precision = 0.99 Commercial comparison Concept drift | Off-device | Using F-score to find most discriminate features | | Provides malice score for apps to better communicate risk, as opposed to binary malware/benign labeling. Hybrid analysis allows for more accurate detection |
| Shabtai et al. (2014) | Dynamic | Integer and boolean vectors | Manual | LR, Decision Tree, SVM, Gaussian Regression, Isotonic Regression | TPR = 0.8 FPR = 0 Accuracy = 0.87 Run-time | Hybrid (feature extraction on-device; training and detection off-device) | None | | Makes novel use of network traffic patterns of apps for malware detection. Reports on a thorough investigation of the CPU/RAM/Storage overhead of the proposed solution for on-device deployment |

**Table 3** (continued)

| | Feature | | | Model | | | |
|---|---|---|---|---|---|---|---|
| | Extraction | Representation | Selection | ML Approach | Evaluation | Deployment | Explanation | Contributions |
| Suarez-Tangil et al. (2017) | Static | Boolean vectors | Mean decrease impurity | Extra Trees | Accuracy = 99.64% Family classification | Off device | None | Novel use of a high variety of features to combat obfuscation. Novel use of feature ranking for dimensionality reduction |
| Bakour and Ünver (2021) | Static | Grayscale Image | Manual | Random Forest, Decision trees, kNN, Ensembles | Accuracy = 0.98 | Off-device | None | Pioneering work in the use of image representation for Android malware detection. A great variety of image feature extraction techniques are used |
| Casolare et al. (2021) | Static | Color Image | Manual | Random Forest, SVM, MLP, CNN | Accuracy = 0.86 Precision = 0.86 Recall = 0.86 | Off-device | None | Combines dynamic analysis with color image representation for Android malware detection |
| Cai et al. (2018) | Dynamic | Integer vector | Manual | Random Forest | Precision = 0.97 Recall = 0.99 F1 = 0.98 RoC Curve AUC = 0.98 Family classification Concept drift | Off-device | None | Makes novel use of ICC Intents for dynamic detection of Android malware. Can handle reflection when detecting API and system calls. Evaluated concept drift |
| Taheri et al. (2020) | Static | Boolean vector | Manual | FNN, ANN, WANN, KMNN | Accuracy = 0.99 FPR = 0.005 AUC = .99 | Off-device | None | Makes novel use of the hamming distance of static binary features for detecting malware. Extensive comparison of the use of different features and ML algorithms |

A few papers, however, do take the temporal relationship of features into account. In such cases, the features are represented as a sequence of values with an specific order, not simply a vector. As section 3b of Fig. 4 depicts, we identified two examples of feature sequences that the surveyed papers have employed:

- **Opcode sequence**: McLaughlin et al. (2017) used as features the sequence of opcodes in the "Classes.dex" file of each APK. For digestion by the ML algorithm, each opcode is eventually one-hot encoded into a boolean vector. However, the order of the calls is preserved.
- **API call sequences**: Karbab et al. (2018) used the sequence of API calls as their feature set. Each API call is then mapped into a vector using an approach called *word2vec*.

Lastly, some papers use more complex structures to represent their features. As section 3c of Fig. 4 shows, these include:

- **Image**: Bakour and Ünver (2021) created grayscale images of Android APKs and then used image classification techniques to detect Android malware. Casolare et al. (2021) took a similar approach, but constructed color images from System-call traces, instead.
- **Strings**: Sahs and Khan (2012) used Strings extracted from APK files directly as features. In order to allow ML algorithms to digest this data, however, they designed a special Kernel for SVM which maps strings to a vector space. Vidal et al. (2018) also used Strings of system calls as features. They, however, used a sequence matching algorithm, instead of ML.
- **Graph**: Gascon et al. (2013) extract the call graph from the DEX file of each APK and use it directly as a feature for malware detection. To prepare the graph for digestion by the ML algorithm of their choice, they use a Neighborhood hash graph approach which converts the graph into an integer vector. This is done in a way that the semantic relationships with-in the graph is preserved. Sahs and Khan (2012) also uses a similar approach to utilize control flow graphs as ML features.

We should also mention that some approaches use techniques that convert the initial feature sets into intermediate ones with specific characteristics, before final digestion. Zhu et al. (2018), for example, use Principal Component Analysis (PCA) which converts input features into a series of orthogonal ones that could reduce dimensionality and improve performance. Another example is Xu et al. (2018), where the authors feed the initial feature set into a series of Long-Short-Term-Memory (LSTM) layers in their neural network, before they are digested by the compute nodes.

**Feature selection**

Once features are represented in an appropriate format, they can be used to train an ML model. However, sometimes the number of features extracted are too high and could slow down the training and/or the detection process. In such cases, approaches are employed to eliminate less important features that do not contribute significantly to the model. This process if often called feature selection.

Our literature review showed that researchers have used two general approaches for feature selection. The first approach is to do it manually, using either domain knowledge or information from a preceding study, as shown in section 4a of Fig. 4. Manual selection is employed by many of the reviewed papers (more than 40%), as demonstrated in Table 3. For example, Liu and Liu (2014), Arp et al. (2014), and Yuan et al. (2016) use domain knowledge to include only permissions or APIs that are known to be associated with malicious behavior. Alternatively, Li et al. (2018) conducted some preliminary studies to compare the distribution of features among malware and benign apps. This allowed them to rank permissions based on the strength of their association with the malware set. And reduce the number of permissions that need to be processed.

The second approach is to use an algorithmic solution, as illustrated in section 4b of Fig. 4. These are techniques that have been used by ML researchers in a variety of research fields. We found the most prevalent solution in the surveyed papers to be "Information Gain" (used by Alzaylaee et al. (2020), Li et al. (2018) and Zarni Aung (2013)). Other techniques employed include "Mutual Information" [used by Wang et al. (2014)], Sequential Selection [used by Wang et al. (2014)], Chi-square [used by Wang et al. (2014) and Feng et al. (2018)], PCA [used by Wang et al. (2014)] and Fischer score [used by Lindorfer et al. (2015)]. Also, some papers (e.g., Li et al. 2018; Zhu et al. 2018) use a combination of these techniques.

**Model creation (training)**

This stage of the ML pipeline involves choosing the algorithm and architecture of the ML model, and then use the selected feature set to train it.

Often, the objective of the model is to label APKs as either malware or benign. Hence, a supervised classification algorithm is used. Some papers, however, aim not only to detect malware, but also to identify the family that the malware belongs to (e.g., whether it is a variant of a known banking trojan). In this case, an unsupervised clustering algorithm may be employed. This distinction is illustrated in section 5 of Fig. 4. Note that in clustering schemes, the ground truth is not used directly in the training of the model. Rather, it is used by the human analysts to label clusters based on the APKs in them.

In case of supervised learning, we observed the surveyed work to use a variety of approaches. As shown in Table 3 and Fig. 4, they comprise three categories:

- **Traditional ML**: These include the ML algorithms that have shallow architectures (i.e., they do not process data in numerous layers, like deep learning does). We observed that the surveyed works have tried a range of these approaches, including logistic regression [used by Amos et al. (2013) and Lindorfer et al. (2015)], tree-based (e.g., Liu and Liu 2014; Shabtai et al. 2014), Naive Bayes (e.g., Zhang et al. 2014; Wu et al. 2012) and SVM (e.g., Arp et al. 2014; Sahs and Khan 2012). Additionally, Demontis et al. (2019) proposed a customized ML approach which is a secured version of SVM.
- **Ensembles**: Ensemble models leverage a combination of base ML models to make predictions. The base models are often traditional ones, such as decision tree or linear regression, that are trained on parts or all of the training data. Our literature review showed that the researchers have tried different ways of creating ensembles. For example, Feng et al. (2018) used an stacking of SVM, decision tree and random forest. Alternatively, Peiravian and Zhu (2013) used a bagging approach with SVM and decision tree.
- **Deep learning**: As deep learning models have shown strong results in other areas of computing, we found researchers to have applied them to Android malware detection as well. Evidently, different architectures of deep learning have been tried. Yuan et al. (2014, 2016) for example, tried Deep Belief Networks (DBN). MultiLayer Perceptron (MLP) was employed by Alzaylaee et al. (2020) and Xu et al. (2018). And, McLaughlin et al. (2017) and Zhang et al. (2018) used Convolutional Neural Networks (CNN).

With regards to clustering, we found it to be not as prominently-used as classification. Notable examples include Wu et al. (2012) and Milosevic et al. (2017) who tried both K-Means and EM for malware family detection. Generally, however, it seems that whereas multi-class classification could be a better solution for family detection, it is rarely employed due to the difficulty of obtaining ground truth for it. Evidently, doing so requires deep manual analysis on a large corpus of malware samples to detect their families, which seems unfeasible for most researchers. Lastly, we should note that we also observed some researchers use clustering algorithms in situ of classification ones. Burguera et al. (2011), for example, used K-Means with $K = 2$ (for two classes: malware and benign) to label APKs based on their neighbors in the clusters they end up in.

**Model evaluation**

Once models are trained, they are usually evaluated to determine if they meet certain expectations. This evaluation can be both in terms of malware detection accuracy, or run-time performance and usability. As section 6 of Fig. 4 shows, our literature review revealed that the surveyed papers have performed two distinct types of evaluation.

The first (and by far the most popular) type consists of classification metrics, which determine how well the model can distinguish between malware and benign. As shown in section 6a of Fig. 4, the metrics reported in this regard include:

- **True positive rate (TPR)**: This is also known as recall or sensitivity. It is defined as the ratio of the number of true positives (i.e., apps that are labeled as malware by the model, and are also actually malware) by the total number of positives (i.e., apps that are actually malware, no matter how they are labeled by the model). A high TPR is desirable for a malware detector because it shows that the solution can correctly detect a greater proportion of malware samples without missing many. Most papers report TPR, such as Liu and Liu (2014) and Arp et al. (2014).
- **True negative rate (TNR)**: Also known as specificity, it is defined as the ratio of the number of true negatives (i.e., apps that are labeled as benign by the model, and are also actually benign) by the number of negatives (i.e., apps that are actually benign, no matter how they are labeled by the model). A high TNR is desirable. It signifies that if the model labels an app as benign, it is very likely actually benign. This can increase users' confidence in the system. TNR is not reported very often, as other metrics cover it. Alzaylaee et al. (2020) is an example of a paper that does report it.
- **False positive rate (FPR)**: It is defined as the ratio of the number of false positives (i.e., apps that are labeled as malware by the model but are actually benign) by the number of negatives. A low FPR is desirable as it signifies that the model rarely mistakes a benign app for a malware. This results in increased usability of the model, as the user does not have to deal with a lot of false warnings. Usually, either TPR or FPR is reported. Arp et al. (2014), for example, reports FPR.
- **False negative rate (FNR)**: It is defined as the ratio of number of false negatives (i.e., apps that are labeled as benign by the model, but actually are malware) by the number of negatives. A low FNR is desirable as it signifies that if the model clears an app as benign, it is very likely benign. Usually, either TNR or FNR is reported. Alzaylaee et al. (2020) m for example, reports FNR.
- **Accuracy**: It is defined as the ratio of the sum of true positives and true negatives by the total number of samples. As it can be seen in Table 3, accuracy is the most popular metric for reporting, as it encapsulates both true positives and negatives. A higher accuracy is desirable as it means the model classifies most examples correctly (either as malware or benign). However, accuracy can also be misleading when there is class imbalance in the data. For example, if the number of benign samples in the dataset is much higher than the number of malware ones, a high accuracy might mask the fact that the model cannot detect many of the malware, simply because the number of true negatives diminishes the effect of true positives.
- **Precision**: It is defined as the ratio of the number of true positives by the sum of the number of both false and true positives. As the name implies, it basically demonstrates how often the labeling provided by the model is correct. A high precision is desirable as it signifies the model does not produce many false results. Examples of papers that report precision include Wang et al. (2016) and Karbab et al. (2018).
- **F-score**: Also known as F1, is provided by a few papers (e.g., Sahs and Khan 2012; McLaughlin et al. 2017). A higher F1 score is desirable as it signifies a better balance between precision and recall.
- **Receiver operating characteristic (ROC) curve**: Some papers (e.g., Wang et al. 2014; Demontis et al. 2019) provide ROC curves, which plots TPR against FPR. It can help users of the system better determine their required threshold. To put simply, ML models usually produce a probability for the maliciousness of each APK, and then label an APK as malware if this probability is above the said threshold. Changing the threshold can allow users to signify whether they have higher tolerances for false positives or negatives. For example, if a user requires very low FPR, they can increase the said threshold until only apps with very high malicious probability are classified as malware. This, however, will also mean that fewer apps with mid-range probability values will be classified as malware, potentially reducing TPR.
- **Area under the curve (AUC)**: A few papers (e.g., Feng et al. 2018; Zhu et al. 2018) provide AUC, which represents the area under the ROC curve. Users can use AUC to compare different models without committing to a specific threshold value.

We should note that a few papers (e.g., Feng et al. 2018; Karbab et al. 2018) also report on classification metrics per malware family. This can help users understand what types of malware the approach could potentially miss, and subsequently devise solutions to mitigate them.

The second type of evaluation conducted by the surveyed papers includes feasibility and usability evaluations. These provides a comprehensive view of the practicality of the proposed approaches. As shown in section 6b of Fig. 4, these evaluations include:

- **Run-time performance**: How quickly can the model digest the training set or label new apps is an important metric, especially when it comes to practical deployment of the model. Yet, we observed that not many of the surveyed papers actually report this data, which hinders judgment of the practicality of their solution. And, even those papers who do provide this information, do it in different and often incompatible ways. Arp et al. (2014), for example, reports on how long it takes their model to digest 1 APK (in this case it is about 1 minutes), whereas McLaughlin et al. (2017) and Zhang et al. (2014) report only the total time that it takes to train the model using all the samples. Feng et al. (2018) report the amount of time their approach takes in each individual step of the training process, from dynamic analysis to feature extraction, feature vector generation, and detection.
- **Resilience to attacks**: A few papers evaluate and report the resiliency of their approach to various attacks, to demonstrate the real-world applicability of it. Demontis et al. (2019), for example, demonstrated how their approach can resists black- and white-box evasion attacks. Alternatively, Kim et al. (2019) showed how their scheme can resists obfuscation attacks, which is often cited as a weakness of static analysis schemes. Lastly, Karbab et al. (2018) investigated how susceptible their approach is to targeted attacks specifically designed to defeat it, by changing the order of API calls.
- **Concept drift (sustainability)**: As Android APIs and malware behavior evolve over time, detection solutions might lose their ability to successfully identify zero-days or new variants. This is often refereed to as concept drift. We observed that very few of the surveyed papers actually investigate how susceptible their approaches are to concept drift. Karbab et al.

(2018) is an exception where the authors investigated how API changes over time might affect the accuracy of their approach. Lindorfer et al. (2015), also, investigated how well can a model trained on older malware (e.g., 2 years in the past) detect newer ones (from the present year). Lastly, Cai et al. (2018) also investigated how their approach would compare to other existing ones, for detecting malware over an 8 years period.

- **Commercial comparisons**: To demonstrate their superiority, some papers (e.g., Lindorfer et al. 2015) report how their approach compares to commercially-available AntiVirus (AV) solutions. This comparison, however, might not provide much insight as not a lot is known about the inner workings of the commercial AVs.
- **Resource overhead**: As we will discuss in the next section, not all Android malware detection solutions can be deployed on device (i.e., on the phone itself). However, for those that can, it is important to measure how they may impact the performance of the phone, and how much resources they may use. These resource can include CPU, RAM or storage capacity that is consumed to extract features, classify samples, store the trained model, or other operations. These metrics are vital to ascertaining the feasibility of any approach. Yet, we observed only a few of the reviewed papers to report them. Dini et al. (2012) were one of such who reported the CPU and RAM overhead of their approach, which were found to be at 7% and 3%, respectively. Shabtai et al. (2014) similarly reported the CPU and RAM overhead of their approach at 1.4% and 13%, respectively. Lastly, when it comes to phones, battery consumption is also essential to consider, as quick battery depletion can easily render a solution infeasible. Surprisingly, we found that only one of the surveyed papers (Saracino et al. 2018) performed this evaluation.

The insights obtained from evaluating the models' resiliency to attacks, in particular, has lead to more research on how to make the ML pipelines more robust. Often refereed to as "Adversarial Robustness," this entails devising approaches that prevents an adversary from feeding the model bogus inputs during training to cause it to mis-classify certain samples (Carlini et al. 2019). This can affect design decisions in any stage of the pipeline, from model training to model explanation.

To this end, Melis et al. (2022) explored whether gradient-based attribution methods, which are used to explain classifiers' decisions by identifying most relevant features, can be used to identify and select more robust learning algorithms for a pipeline. They found that there is indeed a strong connection between uniformity of explanations (entailing that the algorithm has put excess emphasis on a few features) and adversarial robustness. Hence, diversifying feature weights in ML pipelines might be a promising avenue for achieving better robustness.

Rathore et al. (2021) explored robustness of eight different Android malware detection models (based on both traditional machine learning and deep learning) against adversarial attacks. They created the adversarial examples using reinforcement learning. Using their technique, they were able to cause the models to mis-classify up to 86.09% of the samples. Subsequently, they propose an approach based on "Q-tables" to improve the robustness of ML pipelines.

Other similar break-and-fix studies of various Android malware detection models, such as the ones based on Heterogeneous-Graph-based models (Hou et al. 2019) or Image-based classification (Darwaish et al. 2021) has been performed as well. Such insights can inform the design in the model training stage of the pipelines, to make them more robust to adversarial attacks.

**Model use**

This stage covers how the model is deployed to be used by the end-users, and whether there is any attempt to explain either the model or the outcomes of it, to provide clarity to users.

As section 7a of Fig. 4 shows, when it comes to model deployment, the surveyed papers have employed one of the following three methods:

- **On-device**: In this case, the whole ML pipeline is implemented on the phone. The apps are extracted from the device, features are extracted on-device, and the model is trained and used on the phone. Naturally, due to severe computational and resource limitations, very few papers actually attempt this method. We found only Dini et al. (2012) and Saracino et al. (2018) to do so. The advantage of this method, however, is in its self-efficacy and lack of reliance on an external server that might not always be accessible.
- **Off-device**: As it can be seen in Table 3 the majority of the reviewed papers use this model. Here, the entire pipeline is implemented off the phone, usually on a powerful server computer. While this method has the advantage of being less resource-restricted, it is not suitable for end users and is mostly geared towards app market holders.
- **Hybrid**: As a compromise between the two methods above, some papers try to implement parts of the pipeline on-device (usually those that can be done more efficiently on the phone), and offload more computationally intensive tasks (e.g., training) to a remote server. We, however, observed discrepancy between the surveyed work as to how they approach this split. Arp et al. (2014), for example, perform feature extraction and training off-device, while detection and explanation is implemented on-device. In contrast, Li et al. (2018) and Tong and Yan (2017) implement feature extraction on-device, but the rest of the pipeline is offloaded to a server.

In case of explanation, despite its importance for providing clarity to users and establishing trust, we found most of the surveyed works to ignore it. The very few attempted it, used one of the two following approaches:

- **Prediction explanation (interpretability)**: In this case, to explain why an app is labeled as malware/benign by the model, a list of features that contributed the most to the decision is provided. This can help the end user understand the risks associated with installing an app. We found Arp et al. (2014), Gascon et al. (2013), Demontis et al. (2019), and Wang et al. (2014) to be the only ones attempting this approach.
- **Model explanation (explainability)**: In this case, the aim is to help users understand what the model is looking for, when labeling an app as malware/benign. In contrast to the previous approach, here the model as a whole is considered, rather than a specific prediction. We found surveyed papers that attempt this approach to use different techniques. Yuan et al. (2016), for example, use a list of features that have the highest importance values (i.e., are most discriminative between malware and benign). Lindorfer et al. (2015) used a similar approach but used F-score, instead of feature importances. Yang et al. (2014), on the other hand, used the association rule mining technique to present a set of rules based on which apps are labeled as malware/benign by the model.

### Gaps in knowledge and future research directions
In this section, we discuss the gaps that we identified in the reviewed literature.

#### Data collection
While finding benign APKs appears to be straightforward, we found two common issues with sourcing malware APKs, as done by the literature: freshness and accessibility (see "Data collection" section). In terms of the former, our review revealed that most of the popular public datasets are out of date (e.g., MalGenome which was introduced in 2012), which has led many of the proposed approaches to be trained on old data. This issue limits the models' practicality and makes them highly susceptible to concept drift, as we discussed in the previous section.

Accessibility-wise, we found that access to datasets often requires sending a request to their maintainers, which limits the datasets' reach. While it is understandable that access to malware should be restricted to prevent misuse, the way this policy is implemented seems to be problematic. Our investigation of the dataset websites showed that often the person maintaining a dataset is a student who at some point graduates, leaving the dataset without apt maintenance and updating [MalGenome (Zhou and Jiang 2012) is an example of such]. This not only prevents new researchers from accessing the samples, but also makes it challenging to verify the performance of prior models that have used them for training. We should also note that while the prevalence of Advanced Persistent Threats (APTs) is on the rise, there seems to be no datasets that provide a set of Android APT, hindering future research on this import topic.

Clearly, therefore, there is need for better datasets that are more complete, fresh and accessible. This is an important gap that we believe future research should address. As of now, AndroZoo (Allix et al. 2016) seems to be the closest to this ideal. However, it does not provide APTs and still suffers from accessibility issues.

Lastly, we believe another area of improvement to be the way ground truth is obtained/established. Currently, most papers rely on VirusTotal (2021) to label training samples as either malware or benign (as discussed in "Data collection" section). This approach, however, has several drawbacks. Firstly, it might lead to incorrect labels due to incorrect thresholds or putting equal weights on all commercial AV products when they are obviously of different quality. Secondly this approach leads to the trained models mimicking commercial AVs, instead of striving for zero-day detection. In a sense, the model will be trained to be just an aggregator for VirusTotal.

Obviously, manual analysis is the better option for ground truth labeling. However, doing so is often infeasible, due to limited resources. As such, it seems a suitable avenue for future researchers to be providing a carefully labeled (at least verified by manual analysis) set of malware and benign APK (e.g., an updated version of MalGenome) to promote more accurate ML training.

#### Feature extraction
Regarding this stage, our review showed an imbalance in the literature. This is when static features have seen much more attention than dynamic ones (see "Feature extraction" section). This seems to be mostly due to the fact that performing static analysis is usually less computationally expensive. And it also does not require specialized hardware (e.g., real phones to execute the APKs on). The advantage of dynamic analysis, however, is that its more effective against code obfuscation or encryption, as discussed in the previous section. Reports show that the number of obfuscated malware are on the rise (see McAfee 2021), which thwarts static analysis efforts. As such, we believe that future research is needed to explore further the use of dynamic analysis for Android malware detection.

Other than engineering new dynamic features, however, future research can improve other aspects of the analysis, as well. For instance, it is obvious that successful dynamic analysis requires proper input generation to maximize code coverage. Yet, of the dynamic papers we reviewed, only two (Alzaylaee et al. 2020; Wu and Hung 2014) specifically addressed how they generate inputs. Others seem to relied on the default configurations of the tools they used, which might not be optimal. This is an important research oversight that we believe future work should address.

Lastly, we should note that while some of the reviewed papers (Yuan et al. 2014, 2016; Alzaylaee et al. 2020; Saracino et al. 2018; Lindorfer et al. 2015) utilize both static and dynamic analysis, none uses them in tandem. This means that, for example, they do not leverage the data from static analysis to perform better dynamic analysis (e.g., increase code coverage). This is a major oversight by this literature, as this combination has been shown to be highly effective in other domains, such as vulnerability detection (Holland et al. 2016), or web input sanitization (Balzarotti et al. 2008). Thus, we believe statically-informed dynamic analysis for Android malware detection to be a promising avenue of future research.

**Feature representation**

Our review showed that most of the surveyed work put more effort and emphasis on feature engineering. They then opt to use simple representation techniques such as existence-based boolean vectors (see "Feature selection" section). We believe this to be a missed opportunity, as better representation could improve the performance of the models without an increase in processing capacity or engineering complexity (for evidence, see Table 3 where approaches with similar feature sets but more complex representations yield better results). Based on this observation, we believe a needed avenue of future work to be providing insight into how best features can be represented for Android malware detection, or what novel representation techniques can be used (e.g., representing APKs as pictures).

**Feature selection**

We observed that often researchers rely on domain knowledge to eliminate less important features for Android malware detection (see "Feature selection" section). However, we found that the rationale behind these selections are often not discussed with enough detail or justified concretely. For example, while most authors decide to eliminate certain APIs or permissions from analysis [e.g., Arp et al. (2014) eliminates non-"suspicious" API calls, and Yuan et al. (2016) only selects certain permissions], they do not provide evidence (e.g., results of a prior study) that support their eliminations. This makes it difficult to judge their solutions' resilience to concept drift, as the eliminated features might come into usefulness for detection of zero-day malware (evidently, there is already disagreement between the discussed papers as to what counts as "sensitive" permission or API). We believe this to be an important aspect that needs to be paid attention to by future researchers.

Also, when it comes to algorithmic approaches, there is no work (to the best of our knowledge) that investigate and/or compare the fitness of different feature selection algorithms, or provide any insight into the requirements of feature selection for Android malware detection. This seems to lead authors of the surveyed work to simply select the approach that is most familiar to them. Hence, we believe a necessary direction for future research is to perform empirical studies and provide insight into this problem. Our work in this paper, specially Table 3, could be starting point for such research.

**Model creation (training)**

Overall, we found great diversity in the ML algorithms employed by the surveyed works (see "Model creation (training)" section and Table 3). Some papers (e.g., Yerima et al. 2014; Feng et al. 2018) even provided comparisons on the fitness of different ML approaches for Android malware detection. However, we identified a gap in literature to be the use of transfer learning for mobile malware detection. This approach, which uses models trained on differently-distributed data for purposes other than initially-intended, has been frequently deployed and shown promise in a variety of research fields (Pan and Yang 2009). Transfer learning could be specially well-suited for Android malware as it could alleviate the difficulty of obtaining a corpus of malicious APKs.

Another promising area of future research could be time-series learning to analyze historical trends and detect zero-days or new variants of Android malware. While some our surveyed papers leveraged similar approaches, such as Burguera et al. (2011) who employed comparison of different versions of the the same app for detecting re-packaged malware, a more systematic approach to time-series analysis might lead to better alleviation of concept drift.

**Model evaluation**
We found the surveyed works to have performed/reported a plethora of evaluations–from classification metrics, such accuracy and recall, to more complex investigations, such as usability evaluation and malware family detection (see "Model evaluation" section). However, we found that none have conducted all. Worse, we observed that important aspects such as concept drift, resource consumption, and resilience to attacks are often omitted by researchers (see Table 3 for evidence). This is an important research oversight that hinders apt judgment of feasibility and deployability of the schemes. Concept drift (i.e., sustainability of the models over time as malware techniques evolve), in particular, is important aspect. It can help to understand how future schemes can be designed to be longer-lasting. An example of how this evaluation can be done is to train models on older sets and evaluating them using newer ones, as done by Lindorfer et al. (2015). There are other papers, such as (Cai 2018, 2020), and Cai et al. (2020), that investigate the sustainability of Android malware detection through analyzing malware evolution over time. However, these investigations are still not a replacement for an empirical investigation of concept drift in the proposed approaches. Such insight can only be provided by side-by-side comparisons of the approaches using the same dataset, and while keeping the conditions of the study consistent for all. Such a study is still missing from the literature, to the best of our knowledge.

Similarly, investigating resilience of the proposed schemes to different types of attacks is often overlooked, even though it can help inform the design of more flexible malware detection systems. An example of such progress is how Demontis et al. (2019) improved the security of DREBIN [proposed by Arp et al. (2014)] by conducting such security evaluations. Conducting further similar enhancements can be a promising avenue for future research.

Also, for approaches that aim for on-device or hybrid deployment, it seems crucial for researchers to report on resource consumption to demonstrate the feasibility of their approach, as done by Dini et al. (2012) and Shabtai et al. (2014). This will help with better deployability and practicality assessment of the proposed approaches, by other researchers and practitioners.

Lastly, we should note that although hyper-parameter tuning and model calibration are important parts of any ML pipeline, we found the literature to rarely discuss them. Of the papers we reviewed, only Xu et al. (2018) and Karbab et al. (2018) provided information on the former, whereas none discussed the latter. This is an important research oversight, as while this information might not seem fundamental to ascertaining the novelty of the proposed solutions, it hinders others' ability to verify/replicate the reported results. It also limits practical deployment of the solution.

**Model use**
As mentioned before, model use consists of model deployment and model explanation. With regards to the former, we found that the literature usually forgoes this important aspect (see "Model use" section). Often, the focus of the researchers is on providing the best accuracy possible without paying clear consideration to the limitations of the deployment platform. This is not ideal as without a clear path to deployment, any improvement in accuracy is mere untapped potential. This leads us to recommend that future researchers clearly explain how they envision deploying their system (i.e., fully on-device, fully off-device or a hybrid model). They could also provide evidence (e.g., run-time performance data) that clearly demonstrates the feasibility of their deployment plan.

Also, we found the literature to be severely lacking in terms of model explanation, as discussed in the previous section. Most of the reviewed papers did not attempt any form of it (either to explain their model or its predictions), while those who did rarely discussed the actual user experience (e.g., how well the explanations will be perceived and understood by users). Conducting this investigation is important because proper explanation prevents user misunderstandings and potential distrust in the system (Bhatt et al. 2020).

## Reviewed papers
In this section, we provide a timeline and brief summaries of the reviewed papers.

**Timeline of the publications**
Table 4 provide a timeline of the reviewed papers. As it can be seen, they span nearly a decade, from 2011 (which is only 3 years after Android's first release) up to 2020. As the table shows, Android malware detection has been a popular topic of research continually.

**Summary of the papers**
To provide clarity on the approach proposed by each of the reviewed papers, we provide in this section brief summaries of them, in chronological order. A comparison of the papers, on the approaches they use in various stages of their ML pipelines, was presented in Table 3 and was discussed further in "Gaps in knowledge and future research directions".

**Table 4** Timeline of the reviewed papers

| Year | Papers |
| --- | --- |
| 2011 | Burguera et al. (2011) |
| 2012 | Sahs and Khan (2012) |
|      | Wu et al. (2012) |
|      | Dini et al. (2012) |
| 2013 | Aafer et al. (2013) |
|      | Yerima (2013) |
|      | Peiravian and Zhu (2013) |
|      | Gascon et al. (2013) |
|      | Sanz et al. (2013) |
|      | Zarni Aung (2013) |
|      | Amos et al. (2013) |
| 2014 | Arp et al. (2014) |
|      | Liu and Liu (2014) |
|      | Wang et al. (2014) |
|      | Wu and Hung (2014) |
|      | Yerima et al. (2014) |
|      | Yuan et al. (2014) |
|      | Zhang et al. (2014) |
|      | Yang et al. (2014) |
|      | Shabtai et al. (2014) |
| 2015 | Yerima et al. (2015) |
|      | Lindorfer et al. (2015) |
| 2016 | Yuan et al. (2016) |
|      | Wang et al. (2016) |
| 2017 | McLaughlin et al. (2017) |
|      | Milosevic et al. (2017) |
|      | Tong and Yan (2017) |
|      | Suarez-Tangil et al. (2017) |
| 2018 | Feng et al. (2018) |
|      | Karbab et al. (2018) |
|      | Li et al. (2018) |
|      | Xu et al. (2018) |
|      | Zhang et al. (2018) |
|      | Zhu et al. (2018) |
|      | Saracino et al. (2018) |
| 2019 | Demontis et al. (2019) |
|      | Kim et al. (2019) |
|      | Cai et al. (2018) |
| 2020 | Alzaylaee et al. (2020) |
|      | Taheri et al. (2020) |
| 2021 | Bakour and Ünver (2021) |
|      | Casolare et al. (2021) |

Burguera et al. (2011) proposed Crowdroid, a pioneering approach, which compares the execution traces of different versions of the same app, to detect repackaged malware.

Sahs and Khan (2012) use as features list of requested permissions from the Manifest file, and also a control flow graph from the DEX file.

Wu et al. (2012) proposed DroidMat, which uses as features requested Permissions, component names, Intents and API calls. It then performs clustering on the samples, using the K-means algorithm. The number of clusters (i.e., K) is determined by Singular Value Decomposition (SVD).

Dini et al. (2012) proposed MADAM. It extracts dynamic features from two levels, system calls in the kernel-level, and SMSes sent and the user presence status in the user level. They used the kNN algorithm (with k=1) to distinguish between malware and benign.

Aafer et al. (2013) proposed DroidAPIMiner. It uses API call information for Android malware detection. The scheme performs static analysis to extract as features the Android package name of each API call in the disassembled DEX file, and the Permissions requested in the Manifest file of each app. The authors perform manual feature selection where they remove from the model all API calls that were exclusively called by third-party packages.

Yerima (2013) proposed an Android malware detection approach using Bayesian classification. It performs broad static analysis on the app samples to extract features, including API calls, Linux system command strings, requested permissions, encryption routines and presence of secondary APK files. Mutual information is used to eliminate non-important features.

Peiravian and Zhu (2013) proposed an scheme that uses permissions and API calls for Android malware detection. The scheme statically extracts the requested permissions from the Manifest and the API calls from the DEX file of each APK. The features are represented as boolean vectors. The authors tried SVM, Decision Trees and Bagging ensembles as their ML approaches.

Gascon et al. (2013) proposed one of the very first approaches to use API call graphs for Android malware detection. They proposed a novel approach for extracting the call graphs for DEX. They also made novel use of neighborhood hash graph kernel approach to embed the graphs into integer vectors. An SVM model was trained on these vectors.

Sanz et al. (2013) proposed PUMA, an approach that utilizes permission usage for Android malware detection. To use as features, it statically extracts requested permissions and hardware declarations from Manifest files. Features are represented as boolean vectors and are used to train a variety of models, from Logistic Regression, to BayesNet and Random Forest.

Zarni Aung (2013) proposed another similar approach for permission-based Android malware detection. Information gain was used to select only permissions that are most useful for distinguishing between malware and benign. A variety of models was trained on the feature set, including KMeans, Decision Tree, and Random Forest.

Amos et al. (2013) proposed an approach for applying machine learning classifiers to dynamic Android malware detection at scale. It dynamically gathers the following information about each sample: battery usage, binder interactions, memory and network usage, and permission usage. These information are then embedded into boolean vectors and used to train a variety of models (including Random Forest, Naive Bayes, MLP, BayesNet, Logistic Regression and Decision Tree). No specific feature selection was reported.

Arp et al. (2014) proposed DREBIN, a lightweight and explainable Android malware detection method. DREBIN performs static analysis on apps to extract several features including hardware components and permissions declared in the Manifest file of the app, and restricted and suspicious API calls in the disassembled DEX file of it. These feature are then used to train a liner SVM classifier to discern malicious and benign apps.

Liu and Liu (2014) proposed a two-layer Permission-based detection scheme for Android malware. In the first layer, they use the Permissions declared in the Manifest file of each app as features and perform light weight classification on samples. If an app cannot be successfully classified in the first layer, it will be passed to the second layer where the authors perform static analysis to find Permissions that have actually been used by the app (not just simply declared in the manifest file). The second layer is more computationally expensive but also more accurate. For both layers, they use Decision Tree classifiers and use a training dataset of around 28000 benign apps they downloaded from AppChina and around 1500 malware samples they obtained from the Malware Genome Project (MalGenome) (Zhou and Jiang 2012).

Wang et al. (2014) proposed a similar permission-based approach to Li et al. (Wang et al. 2014). They first assess and rank each Android permission (or pairs of Permissions) based on the "risk" it introduces to Android. They then use the "riskiest" Permissions as features to train different classifiers (SVM, Decision Tree and Random Forest) for malware detection. The risk for each permission is determined by how strongly each is associated with (i.e., how frequently it is used by) each class (malware of benign). The authors tried different methods for ranking Permissions, including Mutual Information, Pearson Correlation Coefficient, and T-test.

Wu and Hung (2014) proposed DroidDolphin, one of the first schemes to combine machine learning with dynamic analysis for Android malware detection. The dynamic features extracted include: API calls, network and file operations, started services, loaded classes, information leaks, and actions taken (e.g., SMS sent). An n-gram representation of these features are used to train an SVM model for malware-benign classification.

Yerima et al. (2015) proposed an approach that combines the efficiency of Ensemble learning (specifically, Random Forest) with static analysis. The features used include API calls, Linux command strings, and Permissions. Mutual Information is used for feature selection.

Yuan et al. (2016) proposed DroidDetector, an scheme that uses deep learning, in tandem with both static and dynamic analysis, to detect Android malware. Static analysis is performed to extract permissions and sensitive APIs calls. Dynamic analysis is used to detect executed app actions, such as sending SMS. Features were extracted from a corpus of 20,000 benign and 1760 malware apps obtained from Google play, MalGenome and Contagio Mobile Dump (Contagio 2021). The authors used a Deep Belief Network (DBN) model.

Zhang et al. (2014) proposed DroidSIFT, a scheme that uses weighted contextual API dependency graphs to create a semantically detect Android malware. The authors trained a Naive Bayes model on 13,500 benign apps from McAfee and Google play store, and 2200 malware apps from MalGenome and McAfee. The features in the model are the similarity scores between the dependency graph of the apps in the graph. A manual approach, based on expert-knowledge, was used to eliminate features and improve performance.

Yang et al. (2014) proposed DroidMiner. It statically extracts, from the DEX file of each APK, two types of graphs: one component dependency graph and separate API call graphs for each component. They are then combined to form one that represents the overall behavior of the app. The authors used different paths of this graph as "modalities" to represent specific malicious or benign behaviors. They analyzed a large corpus of apps and gathered a set of modalities. A machine learning approach was then developed based on these modalities. The feature vectors were boolean vectors that indicate whether each of the known modalities are included in the given app. Different ML approaches were tried, including Naive Bayes, SVM, Decision Tree and Random Forest.

Shabtai et al. (2014) proposed an scheme that utilizes deviations in application network behavior to detect malware. It performs dynamic analysis on apps and extract the following information at specific intervals: sent/received data in bytes and percentage, phone's network state (Cellular or WiFi), time since last sent/received data, send/receive mode (if the last transmit event happened a certain time ago). They manually engineer features from these data (e.g., maximum, minimum and average) to train a variety of models, including Logistic Regression, Decision Tree, SVM, and Gaussian Regression.

Lindorfer et al. (2015) proposed MARVIN, a hybrid scheme that combines static and dynamic analysis with ML. MARVIN extracts a large variety of features. Statically, it extracts Java package name, permissions, filtered intents, and publisher id associated with the advertisement libraries. From the DEX file, it also extracts used permissions, use of Java reflection, cryptographic APIs, and dynamic code loading. From the certificate used to sign the file, it extracts the fingerprint, serial number and whether it is self-signed. From other parts of the APK, it extracts the presence of native libraries, native executable or other suspicions files or scripts. Dynamically, it records file and network operations, phone calling events, data leaks, dynamic code loading and registering broadcast receivers. These features are represented as boolean vectors. Fisher score is used to select the most useful set of features, for training SVM and Logistic Regression models.

Yuan et al. (2014) proposed Droid-Sec, one of the first schemes that combines Deep learning with hybrid analysis (i.e., combination of static and dynamic analysis) for Android malware detection. The features extracted through static analysis include requested permissions and sensitive API calls, while the dynamic analysis features include those logged by DroidBox (Project D 2021a) including Network and File operations. A Deep Belief Network (DBN) was trained on the features extracted from 250 malicious apps obtained from Contagio (Contagio 2021) and 250 benign samples from Google play.

Wang et al. (2016) proposed DroidDeepLearner, an scheme that combines deep learning (specifically, Deep Belief Networks) with static analysis for Android malware detection. Permissions (extracted from Manifest files) and API calls (from decompiled DEX files) are used as features.

McLaughlin et al. (2017) proposed a system that uses a deep convolutional neural network (CNN) for malware detection. It uses raw opcodes of apps as input features. The authors treat the sequence of opcodes of each app as a string that requires analysis (similar to natural language processing). This elimination of the need for manual feature engineering is one of the main contribution of this work.

Milosevic et al. (2017) proposed several machine learning-aided Android malware classification schemes. They perform static analysis to extract features, including permissions. Bag-of-words representations of DEX files were also used as features (created by decompiling the DEX files, concatenating all Java code and then counting the frequency of each Java keyword).Various classification models were trained, using SVM, Naive Bayes, Decision Trees, JRIP and AdaBoost. Eventually, an ensemble of these classifiers were used for final detection.

Tong and Yan (2017) proposed an approach that performs dynamic analysis to extract sequences of system calls. It then uses pattern recognition for malware detection.

Suarez-Tangil et al. (2017) proposed DroidSieve, an Android malware classifier based on static and obfuscation-invarient features. They extracted, from each APK, two types of features. The first was static features which include the existence of API calls related to evasion (e.g., reflection and cryptographic ones), permissions and filtered interns, names of app components, and list of API calls and strings. The second type was the resource-centric features, which included data from the signing certificate of the app, package name, embedded APIs and packages, file mis-matches and assets, and lastly the presence of any native code. For feature selection, they rank the features using the Extra Tree algorithm and according to the *mean decrease impurity* measure and select the top 30 to 40% of them. Using the Extra Trees approach for learning, they evaluated the classification accuracy of their model using a corpus of more than 120k apps.

Feng et al. (2018) proposed EnDroid, a system that combines dynamic analysis with ensemble learning. It uses features from DroidBox (Project D 2021a), which include cryptographic, network and file operations. The authors employ the Chi-square feature selection algorithm to remove noisy and uninformative features. For classifier, they use an stacking ensemble of SVM, Decision Trees, Extra Trees, Random Forest and Boosted Trees.

Karbab et al. (2018) proposed MalDozer, a framework for utilizing Deep Learning (specifically, CNN) with sequences of API calls as features. The sequences are converted to vectors, using word2vec (Mikolov et al. 2013) before digestion by the CNN.

Li et al. (2018) proposed SigPID, a system that performs permission usage analysis to identify "significant" ones that are useful for distinguishing between malware and benign APKs. They used a Multi-Level Data Pruning (MLDP) approach to rank permissions by how often they are requested by malicious or benign apps. They then selected ones that were most associated with one class. They also used Sequential Forward Selection (SFS) to further reduce the number of features. An SVM classifier was trained based on 310,926 benign apps Google play and 5494 malware samples from an unidentified source.

Xu et al. (2018) proposed DeepRefiner, a two-layered Android malware detection using Deep learning and static analysis. In the first layer, the contents of all XML files in each app were concatenated and used as input features for an Multilayer Perceptron (MLP) Neural Network. For samples that could not be conclusively labeled by the first layer, a second layer was introduced which extracted Bytecode semantics from DEX files and fed them to Long Short Term Memory (LSTM) layers in a Neural Network. The authors used a dataset consisting of 38,704 samples from VirusShare and MassVet (Chen et al. 2015) and 47,525 benign apps from Google play, to train and test their model.

Zhang et al. (2018) proposed DeepClassifyDroid, a scheme that leverages a CNN trained on static features,

including permissions, Intent filters, API calls, and constant strings.

Zhu et al. (2018) proposed DroidDet, a system that uses an Ensemble Rotation Forest classifier with static features, including permissions, Intent filters, sensitive API and URLs, and permission rates (defined as the number of permissions requested divided by the size of each Smali file). TF-IDF and cosine similarity were used for feature selection.

Saracino et al. (2018) proposed an extended version of MADAM [originally, by Dini et al. (2012)] with an expanded feature set. They added features based on the metadata from apps' market info, including rating, market name, developer, and number of downloads. They also expanded the dynamic analysis to include SMSes sent to numbers not in the contact list and also more indicators of user presence. The kNN algorithm was unchanged, as was the feature selection approach.

Demontis et al. (2019) proposed a modification to DREBIN (Arp et al. 2014), making it resilient to evasion attacks. They systematically analyzed the possibility of different types of such attacks (categorized by the attacker's goal, knowledge and capabilities). Then, they propose Sec-SVM, a modified version of the SVM algorithm which bounds the weight that can be assigned to each feature to specific boundaries. The rationale was that the bounding would make it difficult for DREBIN to put too much emphasis on any one feature that could lead to evasion.

Kim et al. (2019) proposed a multi-modal deep learning method for Android malware detection. It extracts a variety of features through static analysis, including strings, Opcode frequencies, API call frequencies, requested permissions and environmental features in the Manifest file. Topological Data Analysis (TDA) was used for feature selection.

Cai et al. (2018) proposed DroidCat, a new app classification technique which enhances the state-of-the-art in dynamic feature extraction for Android malware detection. Unlike prior approaches, DroidCat features does not solely focus on API or system calls. Rather, inter-component communication Intents and app resources are also included. Additionally, the approach can also handle reflection when detecting system-calls.

Alzaylaee et al. (2020) proposed DL-Droid, an scheme that uses deep learning in tandem with dynamic analysis. Particularly, state-ful input generation was used to increase the code coverage of the dynamic analysis. The extracted features included API calls, intents and DroidBox features. Information Gain was used to select 12 features for the final model, which was trained on a dataset of 19,620 benign apps provided by Intel security and 11,505 malware samples obtained through various sources.

Taheri et al. (2020) proposed an approach for using Hamming distance of static binary features for detecting Android malware. The authors compared three different APK features, namely Permissions, API calls, and Intents, and four different algorithms, namely First Nearest Neighbors (FNN), All Nearest Neighbors (ANN), Weighted All Nearest Neighbors (WANN), and K-Medoid based Nearest Neighbors (KMNN) in terms of distinguishing between malware and benign. Using all the samples in DREBIN, MalGenome and Contagio datasets, they evaluated the accuracy, FPR and AUC of different combinations of features and algorithms.

Bakour and Ünver (2021) proposed VisDroid, a generic image-based classifier for Android malware family detection. The authors created grayscale images from over 24,000 Android samples. They then extracted two types of image-based features (e.g., Scale-Invariant Feature Transform and Color Histogram) and used them to train six different classifiers, including random forest and bagging ensembles.

Taking a similar approach, Casolare et al. (2021) proposed an approach that creates color images from system-call traces of APKs (obtained through dynamic analysis), and then uses image classification techniques to distinguish between malware and benign. The features they extracted from the images included gradient information, frequency of patters, and color information, such as autocolor correlogram. The authors trained a variety of models on the extracted features, including random forest, SVM, MLP and CNN.

**Related work**

There has been numerous survey papers published on Android malware. However, we observed that none covers all stages of an ML pipeline. Zhou and Jiang (2012), for example, provided one of the first investigations of the different types of Android malware. They manually analyzed a large corpus of malware samples and categorized them based on how they infect devices (i.e., installation methods), how they get activated (e.g., through an SMS from the attacker), and what their malicious payloads are. They also introduced the MalGenome dataset which is used by a number of our surveyed papers. They, however, did not specifically discuss the use of ML for malware detection.

Naway and Li (2018) provided a review of the use of deep learning in Android malware detection. They introduce a detailed categorization of approaches proposed by the literature, in terms of types of features used, datasets used, machine learning approaches, performance results, and place of analysis (i.e., on-device or off-device).

Despite their comprehensive taxonomy, however, their work is limited to deep learning only. It also does not follow the structure of a typical machine learning pipeline, and does not provide information on all its stages. For example, the authors do not discuss feature selection in-depth.

Narudin et al. (2016) investigated the evolution of Android malware and Android analysis techniques. They provided an overview of the static and dynamic analysis techniques used by the literature. For static analysis, they identified the focus to be on Permissions, Intents, Hardware components, and DEX files. They categorized dynamic analysis techniques into three groups: In-the-box, Out-of-the-box, and Virtualization. However, the authors only focused on the analysis techniques. As such, they did not discuss the approaches used in different ML pipeline stages, such as feature selection or model evaluation.

Arshad et al. (2016) surveyed Android malware detection and protection. They provided a comprehensive taxonomy of malware types, evasion techniques, and detection approaches. The malware types they identified included: Trojans, Backdoors, Worms, Ransomwares, and Riskwares. The evasion techniques included Repackaging, Drive-by-download, Dynamic payloads, and Stealth. They categorized the detection techniques into static and dynamic. The static ones included signature-based, permission-based and Dalvik-byte-code-based. The dynamic ones included anomaly detection, taint analysis, and emulation-based. The authors, however, did not specifically discuss the use of ML for malware detection.

Ye et al. (2017) reported on their survey on malware detection using data mining techniques. They provided categorizations for different types of malware, concealment techniques, detection approaches, static and dynamic analysis methods, and classification approaches. The detection techniques identified, for example, included signature-based, pattern-based and cloud-based. Through this categorization, the authors provided an overview of the state of malware and malware analysis from an industrial point-of-view. They, however, only focused on feature extraction and model creation. They did not discuss other stages of the ML pipeline, such as data collection, feature representation, and model evaluation and use. Souri and Hosseini (2018) reported a similar survey with the similar benefits and shortcomings.

Feizollah et al. (2015) reviewed feature selection in mobile malware detection. They categorized feature used by the literature into four categories: static, dynamic, hybrid, and metadata. The metadata features they identified, for example, included requested permissions on apps' store listings, app descriptions and developer information. They identified that their surveyed papers had used two general approaches for feature selection: based on rationalizing (i.e., domain knowledge), and based on ranking algorithms. This aligns well with the results of our survey paper, as well. However, the authors did not discuss other stages of the pipeline. For example, they did not discuss model evaluation, deployment or explanation.

Faruki et al. (2014) reported a survey of malware penetration and defenses on Android. Similar to Arshad et al. (2016), they provided a taxonomy of different malware types, and their penetration and survival techniques. They also provided a categorization of different malware analysis and detection approaches employed by their reviewed literature, categorizing them into static and dynamic ones. The static approaches, for example, included signature-based, component-based, permission-based and bytecode-based. The authors also provided a list of tools used to extract features. They, however, did not discuss other stages of ML pipeline, such as feature representation or model use.

Lastly, Yan and Yan (2018) provided a survey specifically on dynamic mobile malware detection. They introduced a taxonomy of the threats, types of features extracted, and criteria for evaluating detection approaches. They, however, did not specifically discuss the use of ML or any static analysis techniques.

Unlike the works described above, our survey provides insight into all stages of the ML pipeline for Android malware detection. As discussed before, this can alleviate the high barrier to entry to this research field, and make it more accessible to new researchers or practitioners.

**Conclusion**

As Android has become a primary target for malware attacks, researchers have investigated the use of ML for automated malware detection on the platform. Using ML, however, usually requires building a complex multi-staged pipeline. Yet, there had been a lack of comprehensive review of how researchers have approached each stage of this pipeline. This has made it difficult for new researchers to get a grip on the state-of-the-art in this field.

In this paper, we filled this gap by providing a novel procedural taxonomy of ML-based Android malware detection. We discussed how researchers have sourced malicious and benign APKs and what static and dynamic features they have extracted from them. We also explored how the literature have represented this features and eliminated the less informative ones. We categorized what ML algorithms have been used, and how the models have been evaluated and explained.

This review also reveled several shortcomings with the current state-of-the-art, which allowed us to provide suggestions for future work. For example, we observed that dynamic analysis has not seen enough attention from researchers and is rather understudied. We also saw that feature representation is often not paid careful consideration and feature selection is seldom done properly. Lastly, we noticed that promising approaches like transfer learning has not been explored for use in Android malware detection.

**Acknowledgements**
We would like to thank the authors of the reviewed papers for their insightful contributions to this research domain.

**Authors contributions**
Mr. MMK: Reviewed the selected papers for the survey study. Drafted the initial design of the taxonomy. Drafted the text of the manuscript. Dr. IA: Contributed to the design of the taxonomy and drafting of the manuscript. Mr. ADR, Mr. YZ, and Mr. RSG, Mr. HS: Contributed to the design of the taxonomy. Provided feedback on the framing of the manuscript. Provided revisions to the manuscript. All authors read and approved the final manuscript.

**Funding**
The funding for this research has been provided by Huawei Technologies Canada. The funding body have not influenced the design of the study or collection, analysis and interpretation of data.

**Availability of data and materials**
All reviewed papers are either in the public domain or available on the corresponding publishers' websites.

**Declarations**

**Competing interests**
The authors declare that they have no competing interests.

**Author details**
[1]Huawei Technologies Canada Co., Ltd, Vancouver, Canada. [2]Huawei Technologies Canada Co., Ltd, Ottawa, Canada. [3]Simon Fraser University, Burnaby, Vancouver, Canada. [4]Huawei Technologies Canada Co., Ltd, Toronto, Canada.

Received: 24 November 2021   Accepted: 2 March 2022
Published online: 02 August 2022

**Publisher's Note**